\documentclass[%
reprint,twocolumn,jcp,superscriptaddress,showkeys,floatfix]{revtex4-1}

\usepackage{graphicx}
\usepackage{dcolumn}
\usepackage{bm}
\usepackage{amssymb}
\usepackage{amsmath}
\usepackage{color}
\usepackage{ifthen}
\setcitestyle{super}

\newcommand{\etal}{et al.\ }
\newcommand{\new}[1]{{#1}}
\newcommand{\fO}{\mbox{$f_0$}}
\newcommand{\fOc}{\mbox{$f_{0,c}$}}

\newcommand{\figref}[1]{\ref{#1}}
\newcommand{\tabref}[1]{\ref{#1}}
\newcommand{\secref}[1]{Sec.\ \ref{#1}}
\newcommand{\appref}[1]{Appendix \ref{#1}}

\begin{document}

\title{Self-assembly of polymeric particles in Poiseuille flow: A hybrid Lattice Boltzmann / External Potential Dynamics simulation study}

\author{Johannes Heuser}
\email{Johannes-Heuser@gmx.de}
\affiliation{Institut f\"ur Physik, Johannes Gutenberg-Universit\"at Mainz, 
D-55099 Mainz, Germany}

\author{G.\ J.\ Agur Sevink}
\affiliation{Leiden Institute of Chemistry, Leiden University, 
P.O. Box 9502, 2300 RA Leiden, The Netherlands}

\author{Friederike Schmid} 
\affiliation{Institut f\"ur Physik, Johannes Gutenberg-Universit\"at Mainz, 
D-55099 Mainz, Germany}






\begin{abstract}
	
We present a hybrid simulation method which allows one to study the dynamical
evolution of self-assembling (co)polymer solutions in the presence of
hydrodynamic interactions. The method combines an established dynamic density
functional theory for polymers that accounts for the nonlocal character of
chain dynamics at the level of the Rouse model, the external potential dynamics
(EPD) model, with an established Navier-Stokes solver, the Lattice Boltzmann
(LB) method.  We apply the method to study the self-assembly of nanoparticles
and vesicles in two-dimensional copolymer solutions in a typical microchannel
Poiseuille flow profile. The simulations start from fully mixed systems which
are suddenly quenched below the spinodal line. In order to isolate effects
caused by walls, we use a reverse Poiseuille flow geometry with periodic
boundary conditions. We identify three stages of self assembly, i.e., initial
spinodal decomposition, particle nucleation, and particle growth (ripening). We
find that (i) In the presence of shear, the nucleation of droplets is delayed
by an amount roughly proportional to the shear rate, (ii) Shear flow greatly
increases the rates of particle fusions, (iii) in later stages of
self-assembly, stronger shear flows may induce irreversible shape
transformation {\em via} finger formation, in particular in vesicle systems.
The combination of these effects lead to an accumulation of particles close to
the center of the Poiseuille flow profile, and the polymeric matter has
a double peak distribution centered around the flow maximum.

\end{abstract}

\maketitle

\section{Introduction} 

The study of inhomogeneous polymer systems is a central research field in
materials science \cite{fredrickson_book, handbook_boudenne}. In particular,
systems of diblock copolymers have been studied intensely in the last decades
as they exhibit interesting morphologies on the mesoscale
\cite{bates1,foerster_02}.  Theoretical investigations, numerical
studies and computer simulations have led to a good understanding of 
polymer melts of diblock copolymers and their phase behavior \cite{schmid2,
matsen2, matsen1, tyler_morse_05, liu_16, dorfman_16}. 

Specifically, the self-assembly of copolymers in solutions has received
increasing interest in the last years. Depending on the polymer structure in
copolymer systems, mesoscale structures like lamellae, micelles, vesicles or
even more complex structures may assemble \cite{lzhang1,antonietti1, uneyama1,
he2,zhang1}. Such structures occur in nature and have an important
function in cells \cite{seifert1}, and they have a variety of potential
applications in (bio)technology, e.g. for encapsulation and transport
\cite{discher1}.

Modeling the self-assembly of such structures on mesoscales is non-trivial and
requires coarse-graining techniques \cite{langner1}.  Whereas the self-assembly
of single small vesicles from surfactant solutions can be studied by classical
all-atom molecular dynamics (MD) \cite{devries1}, this is not feasible for
larger systems. Simulations of the spontaneous self-assembly of surfactant
vesicles have resorted to coarse-grained MD \cite{marrink1}, Brownian dynamics
(BD) \cite{noguchi3, noguchi4, sevink1}, dynamic Monte Carlo (MC)
\cite{bernardes1, bernardes2, huang1, han1, han2}, dissipative particle
dynamics \cite{yamamoto1} and hybrid MD/multiparticle collision dynamics (MPCD)
\cite{sevink2}.  DPD has also been used to investigate polymersomes
\cite{ortiz1, heli1, xiao1, guo1}, however, these studies were restricted to
rather short polymers. On the other side of the spectrum, field-based methods
such as density functional theory \cite{freed_95, uneyama2} and self consistent
field theory \cite{schmid1,matsen_review} are powerful tools to study
self-organizing inhomogeneous polymer systems. Dynamic extensions of these
field models \cite{fraaije1, maurits1, muller1, honda1, zhang1} have given
useful insight into the mechanisms of nanoparticle self assembly in polymeric
systems \cite{sevink05,sevink07,he2,he3,he1,he4}. However, including fluid flow
and hydrodynamic interactions in such studies has remained a challenge.

A number of studies have considered the behavior of vesicles in external flows
such as Poiseuille flow or shear flow.  They have typically focussed on the
deformations of droplets and vesicles in shear flow \cite{rallison1, kraus1,
guido1, vananroye1, noguchi1, coupier1, doddi1, mendes1} and on lifting forces
that determine their lateral position in the channel \cite{kaoui1, farutin1,
segre1, farutin2, mortazavi1}.  Most of them were based on a particle
description, where the droplets/vesicles are treated as individual (possibly
deformable) objects, and they did not address self-assembly.

Other authors have studied the self-assembly of block copolymer melts under
shear \cite{zvelindovsky1, zvelindovsky2, cui1}. In agreement with recent
dissipative particle dynamics (DPD) simulations \cite{feng1}, they found that
the shear flow induces morphological rearrangements and favors cylindrical
structures that are parallel to the flow \cite{feng1}.  In these studies, the
flow field was imposed and the feedback effect of self-assembly on the
instantaneous flow profile was neglected, i.e., hydrodynamic interactions were
ignored.  However, it is well known that hydrodynamic interactions
can influence structure formation in soft matter quite dramatically
\cite{tanaka_05}, the kinetics is often accelerated and kinetic traps can be
avoided \cite{groot2,noguchi2, zhang1}.  To account for such effects, one must
use a simulation method that couples, in both directions, the field-based
description of the free energy in complex fluids with a hydrodynamic
description of fluid flow in a consistent manner.

One such method was recently developed by Zhang and two of us \cite{zhang1}. 
Unfortunately, the dynamical model did not properly account for the
connectivity of the polymers, i.e., monomers were treated as if they moved
independently from one another. As a result, kinetic pathways of vesicle
self-assembly were observed in simulations that are suppressed in reality.  For
example, block copolymer vesicles in solutions easily merged by fusion in the
simulations \cite{zhang1}, whereas fusions are very rare in reality and in
simulations that use a more realistic, albeit purely diffusive dynamical model
\cite{he1,he2,he3}.


In the present study, we propose a hybrid simulation method that combines a
nonlocal model for the polymer diffusion, the so-called external potential
dynamics (EPD) model \cite{maurits1}, with the Lattice Boltzmann (LB) method
for fluid dynamics \cite{succi1}.  With this new method we investigate the
assembly of mesoscale structures, namely internally structured droplets
and vesicles, in a shear flow in the form of two opposite Poiseuille flows.  

At equilibrium, several pathways of vesicle formation have been reported
from simulations \cite{yamamoto1, he1, he2, xiao1, han1} and experiments
\cite{weiss1, adams1, han1, gummel1, qian1}. Here, we focus on the
nucleation-and-growth pathway which has been observed at low polymer
concentration in field-based simulations \cite{he1, he2}, DPD \cite{xiao1} and
experiments \cite{adams1, gummel1}. We study the effect of Poiseuille flow on
nucleation and ripening of the droplets and vesicles, and
on the distribution of polymeric matter across the channel.

By using EPD, which approximates non-local Rouse dynamics (chains move as a
whole), we expect to obtain more realistic results for the kinetic pathways of
self-assembly, which allows us to identify relevant metastable states.  Such
transient states could be stabilized, e.g., by crosslinking ''on the fly''.
Hence simulations based on the new method not only give insights into the
mechanisms of self-assembly under conditions far from equilibrium, but may also
help to design experimental strategies for making novel types of nanoparticles
which do not correspond to stable equilibrium structures. 

The rest of the paper is organized as follows.  In \secref{sec:model}, we
explain the phyiscal background of our new method and how it is implemented.
Then we present results of the assembly of droplets and vesicles in
polymer solutions in a closed system in \secref{sec:no_external_flow} and
in Poiseuille flow in \secref{sec:shear_flow}, focussing on nucleation in
\secref{sec:nucleation} and on ripening in \secref{sec:ripening}. 
Our results are summarized in \secref{sec:conclusions}.

\section{Simulation Model and Method}\label{sec:model}

We consider diblock A:B copolymers with A-fraction $q_A$, immersed
in an explicit solvent S. The copolymers are modeled by Gaussian chains.

The system is treated within polymer density functional theory, hence the free
energy is written as a functional of the dimensionless local \new{composition}
fields $\Phi_I(\mathbf{r})$ of species $I$ with $I=A,B,S$ for A-monomers,
B-monomers, and solvent particles, respectively. The \new{composition} fields
$\Phi_I$ are normalized such that the actual number density is given by
$\Phi_I/v$ with the average particle volume $v$ (which is taken to be identical
for monomers A,B and solvent). The free energy is then written as ${\cal
F}[\Phi] = F[\Phi]/(\beta v)$ ($\beta = 1/k_B T$ is the Boltzmann constant)
with \cite{schmid1, he1}
\begin{align}
	\nonumber F[\Phi] & = -\bar{\Phi}_S V \ln{(Q_S/V \bar{\Phi}_S)} 
           - \frac{\bar{\Phi}_P}{N} V\ln{(Q_P N/V \bar{\Phi}_P)} \\ 
	&+ \int \mathrm{d} \mathbf{r} \:
           \Big[ \: \frac{1}{2}\sum_{I,J \neq I}\chi_{IJ} \Phi_I \Phi_J 
        \label{eq:free_energy}\\  
	\nonumber &- \sum_I \omega_I \Phi_I 
           + \frac{\kappa_H}{2}(\sum_I \Phi_I - 1)^2 \Big].
\end{align}

Here, $N$ denotes the number of segments per chain, $\bar{\Phi}_P$ and
$\bar{\Phi}_S$ are the average volume fractions of polymer and solvent, $Q_P$
and $Q_S$ are the partition function for a single polymer chain and the
solvent, and the $\omega_I$ represent auxiliary ''potential'' fields which
would generate the same \new{composition} fields $\{\Phi_J (\mathbf{r})\}$ in a
reference system of noninteracting polymers. \new{The last term in Eq.\
(\ref{eq:free_energy}) ensures that $\sum_I \Phi_I \approx 1$ everywhere.  For
reasons of numerical stability we allow for small deviations and introduce a
finite Helfand parameter $\kappa_H$.} The most important control parameters 
are the Flory-Huggins parameters $\chi_{IJ}$, which control the interaction 
strength between different species. 

The \new{composition} fields $\Phi_I$ propagate according to a convection
diffusion equation
\begin{equation}\label{eq:conv_diff}
	\partial_t \Phi_I = -\nabla \cdot \textbf{j}_{I}^D
        - \nabla \cdot (\textbf{v} \; \Phi_I)
    =: (\partial_t \Phi_I)^D + (\partial_t \Phi_I)^C
\end{equation}
(superscripts $D,C$ refer to the diffusion and convection, respectively),
where $\textbf{v}$ is the velocity of the hydrodynamic flow field. We take
the flow field to be the same for all components (in contrast to two-fluid
models such as Ref.\ \cite{doi1}). For the first term, the diffusive part, we
adopt an adiabatic approximation, according to which the characteristic time
scales of internal chain relaxations are much smaller than the relevant
diffusive time scales of the system. Hence chains are taken to diffuse as a
whole and the diffusive current $\textbf{j}_I^D$ has the form
(originally derived by Maurits \etal)\cite{maurits1}:

\begin{equation}
   \label{eq:diffusive_current}
	\textbf{j}_I^D(\textbf{r})
        = - D_I \sum_J 
           \int \mathrm{d}\textbf{r}' P_{IJ}(\textbf{r},\textbf{r}') 
           \nabla_{\textbf{r}'} \frac{\delta F}{\delta \Phi_J(\textbf{r}')},
\end{equation} 
Here, $D_I$ denotes the diffusion coefficient of species $I$, 
and the two-body correlator is given by \cite{maurits1}
\begin{equation}
   P_{IJ}(\textbf{r},\textbf{r}') 
        = - \frac{\delta \Phi_I(\textbf{r})}{\delta \omega_J(\textbf{r}')}.
\end{equation}
To avoid the explicit evaluation of the two-body correlator, we adopt the
EPD approximation of Maurits \etal \cite{maurits1} and assume that 
$\nabla_{\textbf{r}} P_{IJ}(\textbf{r},\textbf{r}') 
= - \nabla_{\textbf{r}'} P_{IJ}(\textbf{r},\textbf{r}')$,
which is definitely true in a homogeneous system with 
$P_{IJ}(\textbf{r},\textbf{r}') = P_{IJ}(\textbf{r}-\textbf{r}')$.
Then the diffusive part of Eq.\ (\ref{eq:conv_diff}) (the first term on the
r.h.s.) can conveniently be rewritten as a local equation for the ''external
potentials'' $\omega_I$ \cite{maurits1,he1}. More specifically, we
exploit the fact that there exists a unique relation between $\Phi_J$ 
and $\omega_J$, hence the time evolution equation (\ref{eq:conv_diff})
can equivalently be written as an equation for $\omega_I$, i.e., 
$\partial_t \omega_I = (\partial_t \omega_I)^D + (\partial_t \omega_I)^C$
with
\begin{equation}
\label{eq:diff_w}
	\Big[\partial_t \omega_I (\textbf{r})\Big]^D
        = -D_I \nabla^2 \frac{\delta F}{\delta \Phi_I}.
\end{equation}

In our previous work (Zhang  \etal \cite{zhang1}), we have used a much simpler
Ansatz for the diffusive flow, $\textbf{j}_I^D = - D_I  \Phi_I (\textbf{r})
\nabla \frac{\delta F}{\delta \Phi_I(\textbf{r})}$ (local dynamics). This
describes a situation where monomers move independently from each other. The
present nonlocal model accounts for the fact that monomers in a chain move
cooperatively. It should be noted that in systems containing sharp
interfaces, the EPD approximation may produce artefacts compared to explicit
particle simulations and more sophisticated schemes must be used (S. Qi et al.,
manuscript in preparation). At weak segregations, however, a systematic study
by Reister and M\"uller has showed that the EPD simulation scheme is superior
to the local dynamics scheme and could reproduce the time evolution of structure
factors in reference particle simulations on polymer demixing at a quantitative
level \cite{reister,muller1}. 

In practice, the \new{composition} fields $\Phi$ are calculated from the
auxiliary potentials $\omega$ as follows \cite{schmid1,muller1}:
One introduces partial partition functions $g(\textbf{r},s)$, which
fulfill the modified diffusion equations 
\begin{equation}\label{eq:mod_diff1}
	\partial_s g(\textbf{r},s) 
    = R_G^2 \Delta g(\textbf{r},s) -  N \omega(\textbf{r},s)  g(s,\textbf{r})
\end{equation}
\begin{equation}\label{eq:mod_diff2}
	\partial_s g'(\textbf{r},s) 
   =  R_G^2 \Delta g'(\textbf{r},s) -  N \omega(\textbf{r},s)  g'(s,\textbf{r})
\end{equation}
Here $s \in [0:1]$ parametrizes the position within a chain, the function
$\omega(\textbf{r},s)$ is equal to $\omega_A(\textbf{r})$ for $s < q_A$, and to
$\omega_B(\textbf{r})$ otherwise, and $R_G$ is the gyration radius of one
chain.  We solve these equations numerically using a pseudo-spectral method
introduced by Tzeremes \etal \cite{tzeremes1}
\begin{align}
	g(\textbf{r},s+\mathrm{d}s) &
             = \exp{\left[-\frac{ N \omega(\textbf{r},s
                  +\mathrm{d}s)}{2} \mathrm{d}s \right]} 
                   \exp{\left[R_G^2 \, \Delta \: \mathrm{d}s\right]}  \\ 
	\nonumber &\times \exp{\left[-\frac{N \omega(\textbf{r},s)}{2} 
                       \mathrm{d}s\right]} g(\textbf{r},s).
\end{align}
The $\exp{\left[\Delta \: \mathrm{d}s \right]}$-part is evaluated in Fourier
space (here $\Delta$ is the Laplacian), and the other part is evaluated
in real space.

The polymer densities are calculated by integrating the partial partition
functions over $s$:
\begin{align}
	\Phi_A(\textbf{r}) &= \frac{V \:\bar{\Phi}_P}{Q_P} 
           \int_0^{q_A} 
             \mathrm{d}s \: g(\textbf{r},s) g'(\textbf{r},1-s) \\
	\Phi_B(\textbf{r}) &= \frac{V \:\bar{\Phi}_P}{Q_P} 
           \int_{q_A}^1 
             \mathrm{d}s \: g(\textbf{r},s) g'(\textbf{r},1-s) \\
	\Phi_S(\textbf{r}) &= \frac{V \: \bar{\Phi}_S}{Q_S} 
           \exp{\left[-\omega_S(\textbf{r}) \right]},
\label{eq:phi_s}
\end{align}
where the densities are normalized by the partition functions $Q_P = \int
\mathrm{d}\textbf{r} \: g(\textbf{r},1)$ and $Q_S = \int \mathrm{d}\textbf{r}
\: \exp{\left[- \omega_S(\textbf{r})\right]}$. 

The full convection-diffusion equation (\ref{eq:conv_diff}) is solved by a
simple Euler-forward scheme with alternating convection and diffusion steps.
The convection steps are most conveniently  performed in terms of the
\new{composition} fields $\Phi$, and the diffusion steps in terms of the
auxiliary fields $\omega$ (via Eq.\ (\ref{eq:diff_w})). The calculation of
$\phi$ from $\omega$ via Eqs.\ (\ref{eq:mod_diff1})-(\ref{eq:phi_s}) is
straightforward.  The calculation of $\omega$ from $\Phi$ is done iteratively
with the conjugate gradients method \cite{fletcher1}. 

The fluid dynamics is modeled with a D2Q9 Lattice Boltzmann (LB) environment
\cite{succi1, gross1}, which is based on a set of discrete velocities
$\textbf{c}_i$, and a lattice with lattice sites $\textbf{r}$, populated by a
number of fluid particles $n_i(\textbf{r})$ with velocities $\textbf{c}_i$.
The local mass density of the fluid $\rho$ and the flow velocity
$\textbf{v}$ are then calculated as $\rho = \sum_i n_i$ and $\rho \textbf{v} =
\sum_i n_i \textbf{c}_i$.  The populations $n_i$ are propagated with a
streaming and a collision step according to

\begin{equation}\label{eq:LB_prop}
  n_i(\textbf{r} + \textbf{c}_i \Delta t, t + \Delta t) 
       = n_i(\textbf{r},t) + \Delta_i (\textbf{r},t) .
\end{equation}
In our implementation, we use a multi-relaxation time LB algorithm
\cite{dhumieres1,dunweg1}. Since our dynamical model is a mean-field model
and thermal fluctuations are not included in the convection-diffusion
equation (\ref{eq:conv_diff}), we also do not thermalize the LB modes
for consistency.

The thermodynamically driven diffusive flow of the polymers in solution induces
fluid flow. To account for this effect, the Navier Stokes equations for a
Newtonian fluid have to be extended, either by including an
additional stress term of the form $\nabla \tilde{\sigma}$ or, equivalently,
a corresponding bulk force term $\textbf{f}$.  We choose the second
variant \cite{zhang1} and write the flow equation in the form
\begin{equation}
   \label{eq:navier_stokes}
   \rho \left(\partial_t \textbf{v}
           + \textbf{v} \cdot \nabla \textbf{v} \right) 
       = \textbf{f} - \nabla p + \nabla \sigma' .
\end{equation}
where $\sigma'$ denotes the viscous stress tensor of a Newtonian fluid and $p$
is the pressure. To implement the force coupling in the LB scheme, we follow
Ref.\ \cite{dunweg1} and extend the collision operator by adding a force
contribution
\begin{equation}\label{eq:force_col}
	\Delta_i'' = a^{c_i} \left[\frac{\Delta t}{c_s^2} f_{\alpha} c_{i \alpha} + \frac{\Delta t}{2 c_s^4} \Sigma_{\alpha \beta} \left(c_{i \alpha} c_{i \beta} - c_s^2 \delta_{\alpha \beta} \right) \right]
\end{equation}
with
\begin{align}
	\Sigma_{\alpha \beta} 
         &= \frac{1}{2} \left(1 + \gamma_s \right) 
         \left[\text{v}_{\alpha} f_{\beta} + \text{v}_{\beta} f_{\alpha} 
           - \frac{2}{3} \text{v}_{\gamma} f_{\gamma} \delta_{\alpha \beta} \right] \\
	\nonumber &+ \frac{1}{3} \left(1 + \gamma_b \right) 
          \text{v}_{\gamma} f_{\gamma} \delta_{\alpha \beta},
\end{align}
where the prefactors $a^{c_i}$ are the weight factors of the D2Q9 model 
\cite{succi1}, $c_s = \Delta x/\Delta t \sqrt{3}$ is the speed of sound, 
$\Delta x$ and $\Delta t$ are the lattice constant and the LB time step,
and $\gamma_s, \gamma_b$ are the relaxation parameters 
of the multi-relaxation LB algorithm, which set the shear and bulk viscosity,
$\eta$ and $\eta_B$, {\em via} \cite{dunweg2} 
\begin {equation}
\eta = \frac{\Delta t \rho c_s^2}{2} \frac{1+\gamma_s}{1-\gamma_s}
, \qquad
\eta_B = \frac{\Delta t \rho c_s^2}{d} \frac{1+\gamma_b}{1-\gamma_b},
\end{equation}
where $d$ is the spatial dimension.

The mechanical force density field $\mathbf{f}$ in Eq.\
(\ref{eq:navier_stokes}), which is to be coupled to LB, is transmitted to the
fluid by the monomer segments or solvent particles and should be identical to
the force driving the diffusive monomer currents. However, the adiabatic
approximation, Eq.\ (\ref{eq:diffusive_current}), creates some ambiguity. From
a thermodynamic point of view, the force density should be evaluated directly 
from the free energy according to \cite{maurits2, honda1} 
$\textbf{f}^{\mbox{\tiny TD}} = \sum_I \textbf{f}^{\mbox{\tiny TD}}_I$
with
\begin{equation}
  \label{eq:force_field_0}
\textbf{f}_I^{\mbox{\tiny TD}}(\textbf{r}) = - \frac{1}{\beta v} \Phi_I \: 
\nabla \frac{\delta F}{\delta \Phi_I}. 
\end{equation}
This thermodynamic force drives the local diffusive motion of monomers on very
short time scales. In the adiabatic approximation, however, monomers of a chain
are taken to move together. Rapid internal chain motions are averaged out, and
chains move as a whole, like rigid bodies, driven by the total thermodynamic
force acting on each chain. Comparing Eq.\ (\ref{eq:diffusive_current})
with the relation $\mathbf{j}_I^D/v = \beta D_I \mathbf{f}_I$ between forces 
and currents, one finds that this corresponds to an independent monomer motion 
in an effective force field
$\textbf{f}^{\mbox{\tiny eff}} = \sum_I \textbf{f}^{\mbox{\tiny eff}}_I$
with
\begin{equation}
  \label{eq:force_field}
  \textbf{f}_I^{\mbox{\tiny eff}}(\textbf{r}) 
    = - \frac{1}{\beta v} \sum_J \int \mathrm{d}\textbf{r}' 
    P_{IJ}(\textbf{r},\textbf{r}') 
         \nabla_{\textbf{r}'} \frac{\delta F}{\delta \Phi_J}(\textbf{r}'),
\end{equation}
In a mechanically consistent model, the force entering (\ref{eq:navier_stokes})
should thus be given by the effective force, Eq.\ (\ref{eq:force_field}), which
is the thermodynamic force averaged over the gyration radius. In the present
work, we choose this second, mechanically consistent type of force coupling.
Similarly, we disregard solenoidal part of $\textbf{f}_I(\textbf{r})$ in Eq.\
(\ref{eq:force_field}), since solenoidal contributions to $\mathbf{f}_I
\propto \mathbf{j}_I^D$ have no effect on the convection-diffusion dynamics,
Eq.\ (\ref{eq:conv_diff}).  In \appref{app:f_rotation}, we show that the
force fields $\mathbf{f}_I$ are in fact purely irrotational within the EPD
approximation $\nabla_{\textbf{r}} P_{IJ}(\textbf{r},\textbf{r}') = -
\nabla_{\textbf{r}'} P_{IJ}(\textbf{r},\textbf{r}')$ and in particular in
homogeneous fluids. 

In practice, the force fields are evaluated as follows. We
reconstruct the irrotational part of $\mathbf{f}_I$ from the change
\new{in the composition field} resulting from the diffusion step.
\begin{eqnarray}
 \nonumber
   \big[ \delta \Phi_I(\textbf{r},t)\big]^D &=& 
     \Big[\Phi_I(\textbf{r},t)\!-\!\Phi_I(\textbf{r},t-\delta t) \Big]^D\!\!
\\ &=&
     -v \beta D_I \nabla \cdot \textbf{f}_I(\textbf{r},t) \delta t
\end{eqnarray}
After Fourier transformation ($\textbf{k}$-space), we obtain an explicit
expression for the force
\begin{equation}
   \hat{\textbf{f}}_I(\textbf{k},t) 
     = \frac{1}{\beta v} \frac{\big[\delta \hat{\Phi}_I(\textbf{k},t)\big]^D}
                   {D_I \delta t} \frac{i \textbf{k}}{|\textbf{k}|^2}  
\end{equation}

This completes the formulation of our model. The diffusive dynamics of the
polymer \new{composition} fields is coupled to the fluid flow via the force
$\mathbf{f} = \sum_I \mathbf{f}_I$ in the Navier-Stokes equations, Eq.\
(\ref{eq:navier_stokes}), and the fluid flow is coupled to the polymer dynamics
via the convection term in the convection-diffusion equation, Eq.\
(\ref{eq:conv_diff}). \new{We note that these are the only two couplings
between the composition fields and the flow fields in the model. In particular,
we do not impose a strict relation between the number densities and the mass
density.  Instead, the total number density $\sum_I \Phi_I$ and the mass
density $\rho$ are approximately kept constant by separate compressibility
terms. This represents an approximation which can be applied in fluids that are
roughly incompressible, and where the local mass density is roughly independent
of the local composition.} A flow chart of the simulation algorithm and
additional explanations are given in \appref{sec:flowchart}.

\new{Our mean-field scheme does not include thermal fluctuations. Formally,
they can be included as in Ref.\ \cite{zhang1} by adding Gaussian noise terms
in the convection-diffusion equation (\ref{eq:conv_diff}) and in the
Navier-Stokes equations (\ref{eq:navier_stokes}), which satisfy the
fluctuation-dissipation theorem. The noise in the hydrodynamic equations can be
implemented in the Lattice Boltzmann framework as described in Refs.\
\cite{zhang1,dunweg1}. The thermodynamically consistent implementation of noise
in the convection diffusion equation within the EPD approximation, however, is
not trivial and requires special efforts \cite{muller1}. One can mimick the
disordering effect of thermal noise by adding a simplified noise term that just
guarantees mass conservation, but violates the fluctuation dissipation theorem
\cite{he1}.  However, such an approximation should only be used to study the
mostly deterministic time evolution of a non-equilibrium systems in the
mean-field regime, where the exact structure of the noise does not matter too
much. It is not suitable for studying equilibrium distributions and thermally
driven processes.  }

In the next sections, we apply our method to study particle self-assembly
in closed systems and in shear flow. These calculations are done in two
dimensions. This allows us to cover a large range of shear rates with good
statistics (many independent simulation runs), and to study large systems over
long times in order to investigate flow-induced shape deformations in late
stages of self-assembly. In principle, however, the method is not
restricted to two dimensions. In three dimensions, the D2Q9 LB scheme
must be replaced by a threedimensional scheme such as the D3Q19 
scheme \cite{succi1}.
 
\section{Particle Self-assembly in Closed Systems}
\label{sec:no_external_flow}

We first consider the self-assembly of copolymeric droplets and vesicles in
closed systems without external flows.  We choose the units of length, time,
and mass such that the radius of gyration $R_G$, the diffusion coefficient of
the solvent $D_S$, and the mass of the solvent $m_S$ are unity, which gives the
time unit $\tau = R_G^2/D_S$.  Furthermore, we set the shear and bulk viscosity
to $\eta = \eta_B = 1 R_G^2/\tau \cdot \rho$ and take the masses of particles
to be equal for all species, hence the mass density is $\rho = m_S/v$.  The
parameter combination $1/(\beta v)$  (with the Boltzmann factor $\beta$) would
set the noise level in a simulation that includes fluctuations \cite{zhang1}.
At the mean-field level considered here, where fluctuations are neglected, the
parameters $v$ and $\beta$ do not enter the results. Hence we do not need to
specify them here.  

To isolate the  effect of hydrodynamics, we choose the other model parameters
according to a previous study by He \etal \cite{he1}, i.e., $N=17$ with a
length fraction $q_A=0.882$ of (hydrophobic) A-monomers, $D_{A,B}=D_S/N$,
$\bar{\Phi}_P=0.1$, and the interaction parameters $\kappa_H=1.176$,
$\chi_{AB}=1.05$, $\chi_{AS}=1.20$. Two values of Flory-Huggins parameters
$\chi_{BS}$ are considered, namely $\chi_{BS} = 0.75$, which leads to
systems of droplets, and $\chi_{BS} = -0.15$, which leads to the assembly
of vesicles \cite{he1,he2}. As the droplets in the first case show
structural resemblance to micelles, we will call these particles
micelle-shaped droplets and refer to the first case as ''droplet systems''.
Systems of the second kind will be called ''vesicle systems''.

The grid spacing $\Delta x$ is chosen $R_G/3$ both for the LB part and the
solution of the convection-diffusion equation, and the timestep is set to
$\Delta t = 0.002 \tau$. The time step determines the ''lattice velocity'' $c =
\Delta x/\Delta t$ in the LB system, which is proportional to the speed of
sound $c_s$, and must hence be chosen small enough that the processes of
interest are slow compared to $c$. In our system, we have $R_g/c \approx 0.006
\tau$, which is much shorter than the time scale of diffusion, $R_g^2/D_S = 1
\tau$.

The system size is chosen $50 R_G \times 50 R_G$
with periodic boundary conditions. The simulation runs start from a perfectly
mixed system, which is homogeneous except for a very small random noise
which varies from system to system. Apart from this small initial
inhomogeneity, there is no further source of noise.  Thermal fluctuations are
not included, and the simulation runs are completely deterministic. This
choice of system setup allows us to make a precise comparison of the evolution
of a configuration with and without hydrodynamics, i.e., with and without
coupling to the LB simulation.

\begin{figure}[t]
\centering		
	\includegraphics[scale=1]{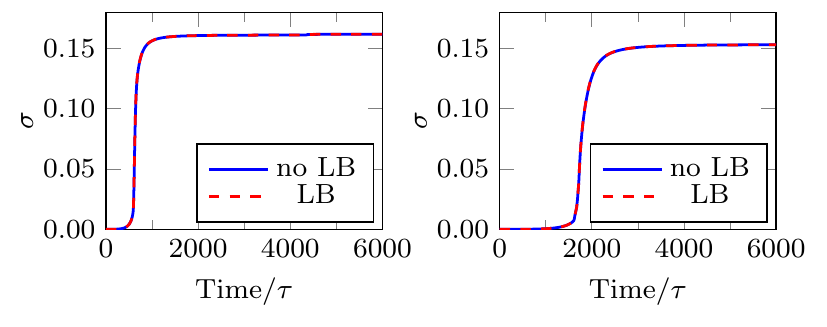}
	\caption{Evolution of $\sigma$ with time $t$ (units $\tau$) in 
a system with hydrodynamics (dashed red curve: EPD convection-diffusion 
equation coupled to LB simulation), and a corresponding system 
(same initial conditions) without hydrodynamics (full blue curve: pure EPD
simulation). Left: droplet system ($\chi_{BS} = 0.75$),
right: vesicle system ($\chi_{BS} = -0.15$)}\label{fig:sigma}	
\end{figure}

Following Ref.\ \cite{he1}, we introduce a quantity which quantifies the 
mixing of solvent and polymers:
\begin{equation}
 \sigma(t) = \frac{1}{V} \int \mathrm{d}\textbf{r} |\Phi_P(\textbf{r},t) 
                  - \bar{\Phi}_P|,
\end{equation}
where $V$ is the volume of the system and $\Phi_P = \Phi_A+\Phi_B$ the
volume fraction of polymers.  The state $\sigma = 0$ corresponds to a
perfectly mixed system, and increasing $\sigma$ signalizes segregation.
\figref{fig:sigma} shows a typical time evolution $\sigma(t)$ in systems set up
as described above. At the beginning, the system decomposes slowly from an
initially almost perfectly mixed state.  The speed of the segregation process
increases until the system reaches a ''nucleation stage'', at which $\sigma$
shows a sudden increase.  Finally $\sigma$ saturates, which denotes the
ripening stage.  In agreement with the results from He \etal \cite{he1}, we
find that nucleation takes place much earlier in systems with $\chi_{BS} =
0.75$, where the final structures are micelle-shaped droplets, compared
to systems with $\chi_{BS}=-0.15$, where vesicles can emerge.

We have also analyzed how the number of particles changes with time after the
initial nucleation stage (data not shown).  Both in the systems with and
without hydrodynamics, it remains roughly constant in the vesicles systems with
$\chi_{BS} = -0.15$, and decreases in the droplet systems with $\chi_{BS}
= 0.75$. In the latter case, large particles are found to grow at the expense
of smaller ones until some of the smaller ones completely dissolved. Fusions
of particles are never observed, neither in systems without nor with
hydrodynamics. 

\begin{figure}[t]
\centering		
\includegraphics[scale=1]{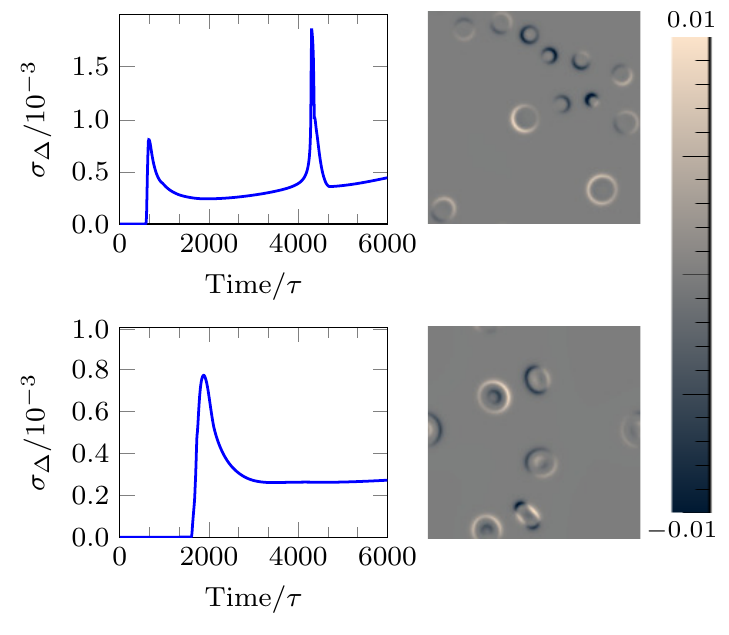}
\caption{
Evolution of $\sigma_{\Delta}$ with time $t$ (left, units $\tau$) in 
droplet systems ($\chi_{BS} = 0.75$, top) and vesicle systems 
($\chi_{BS} = -0.15$, bottom) with corresponding difference snapshots 
of $\Phi_P^{LB}(\textbf{r})-\Phi_P^{NLB}(\textbf{r})$ (right) at 
$t = 2000 \tau$ in the droplet system and $t = 3000 \tau$ in 
the vesicle system. The values of the difference are confined to a
very narrow regime between from $-0.01$ (blue) and $0.01$ (beige).}
\label{fig:sigma_delta}	
\end{figure}

In \figref{fig:sigma}, the curves of $\sigma(t)$ for systems with hydrodynamic
interactions (where the EPD simulation is coupled to a LB simulation) and
without (no coupling) are practically indistinguishable.  To further quantify
the influence of hydrodynamics on self-assembly, \new{we have examined the
polymer structure factor at different times (data not shown). However, in the
time range covered by the simulation, the structure factors of systems with and
without hydrodynamic coupling were practically identical. Hence we introduce
another, more sensitive quantity $\sigma_{\Delta}$, which allows to elucidate
local effects of hydrodynamic flows.} It integrates over the local absolute
difference of the dimensionless polymer distribution $\Phi_P^{LB}(\textbf{r})$
in systems with coupled LB simulation, and the corresponding distribution
$\Phi_P^{NLB}$ in systems without LB coupling (but identical initial
conditions) and is defined as

\begin{equation}
  \sigma_{\Delta} 
    = \frac{1}{V} \int \mathrm{d}\textbf{r} |\Phi_P^{LB}(\textbf{r}) 
                  - \Phi_P^{NLB}(\textbf{r})|.
\end{equation}
Results for an example of a droplet and vesicle system are shown in 
\figref{fig:sigma_delta}. Over all, $\sigma_\Delta$ is very small (of order
$10^{-3}$). It starts at zero and then exhibits a first peak up to around
$0.001$ around the time where nucleation sets in.  Then, $\sigma_{\Delta}$
decreases to values below $0.0005$ and levels off, but may occasionally show
further peaks which correspond to singular events. For example, the second peak
in \figref{fig:sigma_delta}, top, coincides with the dissolution of a
droplet.

The local difference plots on the right side of \figref{fig:sigma_delta}
give further insight into the effect of hydrodynamics on phase separation. In
the droplet system, the larger droplets are surrounded by beige coronae and the
smaller systems by blue coronae, indicating that the growth of large droplets
and the shrinking of smaller droplets is slightly accelerated in the presence
of hydrodynamics.  A more quantitative analysis of difference plots such as
\figref{fig:sigma_delta} showed that fluid dynamics generates a speedup of the
order of one percent. This is much less than in our previous study using local
dynamics \cite{zhang1}, and in a recent study of self-assembly of small
molecules \cite{noguchi2}. Hence, we conclude that hydrodynamic flows have
no significant effect on structure formation in our polymer solutions. The flow
fields generated by the polymer diffusion seem to be too small to have large
feedback effects on the polymer concentration fields. 

As already mentioned earlier, the self-assembly of vesicles in the absence of
hydrodynamic flows had been studied earlier by He \etal \cite{he1} using
nonlocal, cooperative dynamics, and by Zhang \etal \cite{zhang1} using local
dynamics. Our results here agree with those of He \etal \cite{he1}, who found
that fusion of particles is suppressed once the particles have reorganized
themselves internally such that they are surrounded by a hydrophilic corona of
B-monomers. In contrast, Zhang \etal \cite{zhang1} did observe particle fusion
events, hence the corona seems to affect the kinetics only if the monomers move
cooperatively.  

Zhang \etal \cite{zhang1} also studied the effect of hydrodynamic interactions
and found that self-assembly was accelerated particularly in the late stages.
The main effect seemed to be the acceleration of fusion events.
Our present results show that hydrodynamic flows have only a very small effect
on the kinetics of particle assembly, if particle fusion is kinetically
suppressed. 

\section{Particle Self-assembly in Poiseuille Flow}
\label{sec:shear_flow}

Next we investigate the effect of shear flow on the kinetics of self-assembly.
We mimick an experimental situation where micelle-shaped
droplets and vesicles self-assemble in thin channels, e.g., in a microfluidic
device.  However, we are not interested in boundary effects here. To eliminate
them, we follow Refs.\ \cite{backer_05,fedosov_10} and create a system of
opposite Poiseuille flows with periodic boundary conditions (reverse Poiseuille
flow).  This is done by applying a bulk force of the form
\begin{equation}\label{eq:force_field_2}
	f_x(y) = \begin{cases} -\fO \qquad y<L_y/2 
                            \\ +\fO \qquad y \geq L_y/2 \end{cases}
\end{equation}
in $x$-direction to the fluid. The theoretical prediction for the resulting
velocity field (from the Navier Stokes equations) is:
\begin{equation}\label{eq:flow_field}
	v_x(y) = -\frac{\fO}{2 \eta} 
                 \left(y - \frac{L_y}{2} \right) \left( \left|y 
                         - \frac{L_y}{2} \right| - \frac{L_y}{2} \right)
\end{equation}
Here $L_y$ denotes the $y$-size of the system and $\eta$ is the shear
viscosity. 
The resulting shear rate $\dot{\gamma} = \frac{\partial v_x}{\partial y}$ is a
linear function in $y$ with the form
\begin{equation}
	\dot{\gamma} = \frac{\fO}{2 \eta} 
            \left(\frac{L_y}{2} - 2 \left| y 
                     - \frac{L_y}{2} \right| \right)
\end{equation}

In the following, forces will be given in units of $f^* = \rho R_G\tau^{-2}$.
The amplitude of the force field in our simulations ranges from $\fO = 1 \cdot
10^{-5} f^*$ to $\fO = 1 \cdot 10^{-3} f^*$. With these forces, we reach
Reynolds numbers up to $2$ and stay in the regime of low Reynolds numbers.
Moreover, the shear rate is sufficiently small that real polymers with
radius of gyration $R_G$ and diffusion constant $D_P$ would not deform, which
provides a justification for our adiabatic approximation.  (Real polymers with
Rouse time $\tau_R=\frac{2}{\pi^2}\frac{R_G^2}{D_P}$ would have Weissenberg
numbers below $Wi = \tau_R \dot{\gamma} < 0.04$ in our shear flows, and polymer
deformations become important for $Wi > 1.3$ \cite{smith1}.)

We will refer to systems with $\fO < 3 \cdot 10^{-5} f^*$ as ''weakly
sheared'', systems with $3 \cdot 10^{-5} f^* \leq \fO \leq 3 \cdot 10^{-4} f^*$
as ''moderately sheared'', and even higher $\fO$ as ''strongly sheared''.  The
model parameters and the initial simulation setup are the same as in the
previous section. We will again compare ''droplet systems'' with
$\chi_{BS} =0.75$ with ''vesicle systems'' with $\chi_{BS} = -0.15$ and use
simulation boxes of size $50 R_G \times 50 R_G$. For bulk forces up to $\fO = 5
\cdot 10^{-5} f^*$, we average over 100 independent runs per parameter set and
otherwise, over 50 runs.  The systems were initialized with the flow field
defined by Eq.\ (\ref{eq:flow_field}).  Hence the force field just has to
conserve the flow field during the simulation run. 

We will first discuss the initial stage of self-assembly where the first
nuclei appear (nucleation stage, \secref{sec:nucleation}), then analyze
the effect of shear on the ripening stage (\secref{sec:ripening}), examine the
development of the particle shapes (\secref{sec:part_shape}) and the
characteristic relaxation times (\secref{sec:relaxation}), and finally study
the lateral migration of particles and/or polymeric matter across the channel 
(\secref{sec:stdev}).

\subsection{Nucleation Stage}\label{sec:nucleation}

\begin{figure}[t]
\centering		
\includegraphics[scale=1]{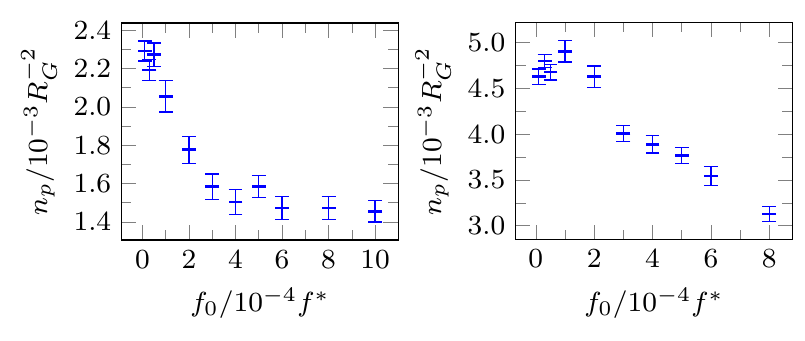}
\caption{
Maximum average particle density during the self-assembly process vs.\ 
bulk driving force $\fO$ in units of $f^*$. 
Left: vesicle systems ($\chi_{BS}= -0.15$),  
right: droplet systems ($\chi_{BS}= 0.75$) }
\label{fig:max_part_density}	
\end{figure}

We begin with investigating the first stage of particle assembly, where
nuclei initially form. We call this stage ''nucleation stage'', even
though the process of self-assembly is deterministic in our system and not
driven by random thermal fluctuations (which are not included in our mean-field
treatment). As discussed in earlier work, the nuclei formation is
triggered by spinodal decomposition in this case \cite{he1,kessler_16}. During
an initial ''incubation time'', spinodal concentration fluctuations build up
until they become large enough that nuclei start to form throughout the
system almost simultaneously. Here we study how the number of these nuclei
depends on the strength of the shear flow. Thus we count the number of
particles right after the nucleation stage, where ''particles'' are defined as
connected clusters of lattice sites with local polymer volume fractions above
$\Phi_P \ge 0.5$.  Specifically, we determine the particle density $n_p$, i.e.,
the average number of particles per system divided by the system size. The
results are shown as a function of $\fO$ in \figref{fig:max_part_density}. 

Small shear flows have little influence on the densities of nuclei. If the
force amplitude $\fO$ exceeds a certain threshold $\fOc$, the particle
densities start to decrease significantly with increasing $\fO$. The threshold
is much smaller in the ''vesicle systems'' with $\chi_{BS}=-0.15$ 
($\fOc \sim 1 \cdot 10^{-4} f^*$) than in the ''droplet systems'' with
$\chi_{BS}=0.75$ ($\fOc \sim 2.5 \cdot 10^{-4} f^*$). Thus, the
nucleation of compact particles made of hydrophobic polymers is less affected
by shear flow than the nucleation of more open particles made of polymers that
also contain strongly hydrophilic blocks.

\begin{figure}[t]
\centering		
\includegraphics[scale=1]{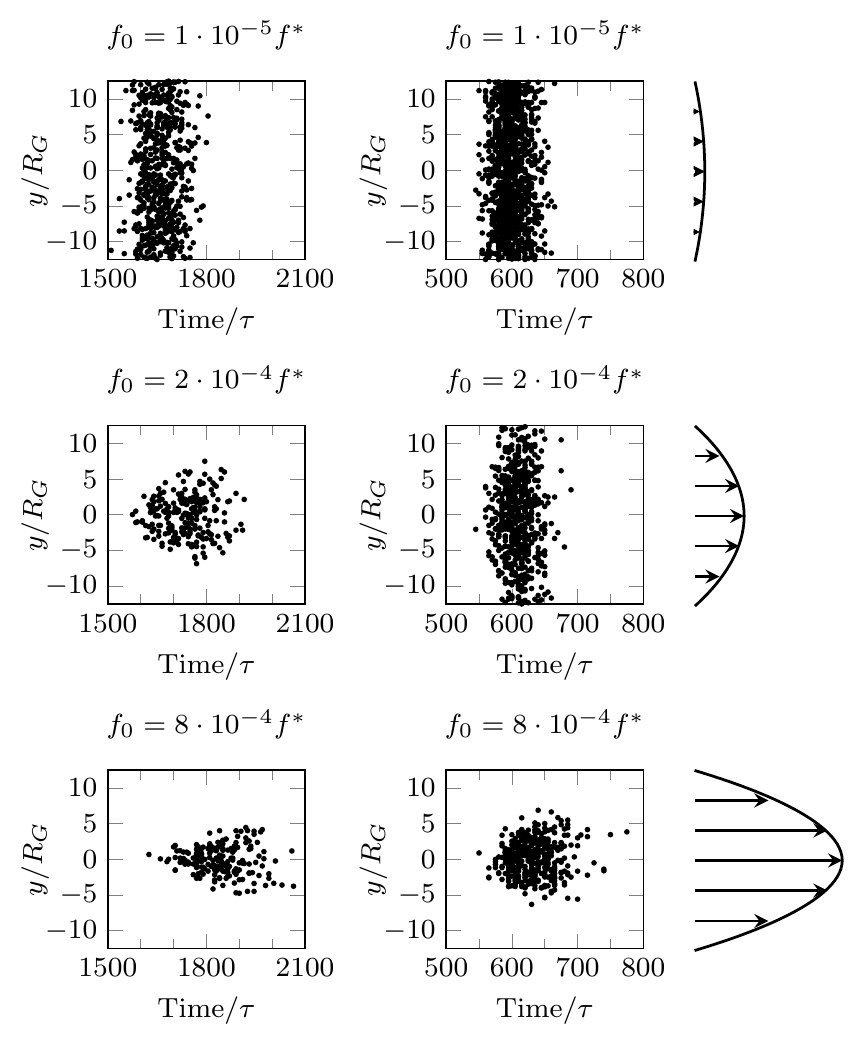}
\caption{Scatter plot of nucleation events (in all simulations) in the
coordinate plane of time vs.\ $y$ ($y=0$ corresponds to the center of the
Poiseuille flow -- see the flow profiles in the outer right column) for
examples of weakly sheared systems (top), moderately sheared systems (middle),
and strongly sheared systems (bottom) with bulk force amplitudes $\fO$ as
indicated.  Left column: vesicle systems ($\chi_{BS}= -0.15$),  right column:
droplet systems ($\chi_{BS}= 0.75$).}
\label{fig:assembly_comp}	
\end{figure}

\begin{figure}[t]
\centering
\includegraphics[scale=.5]{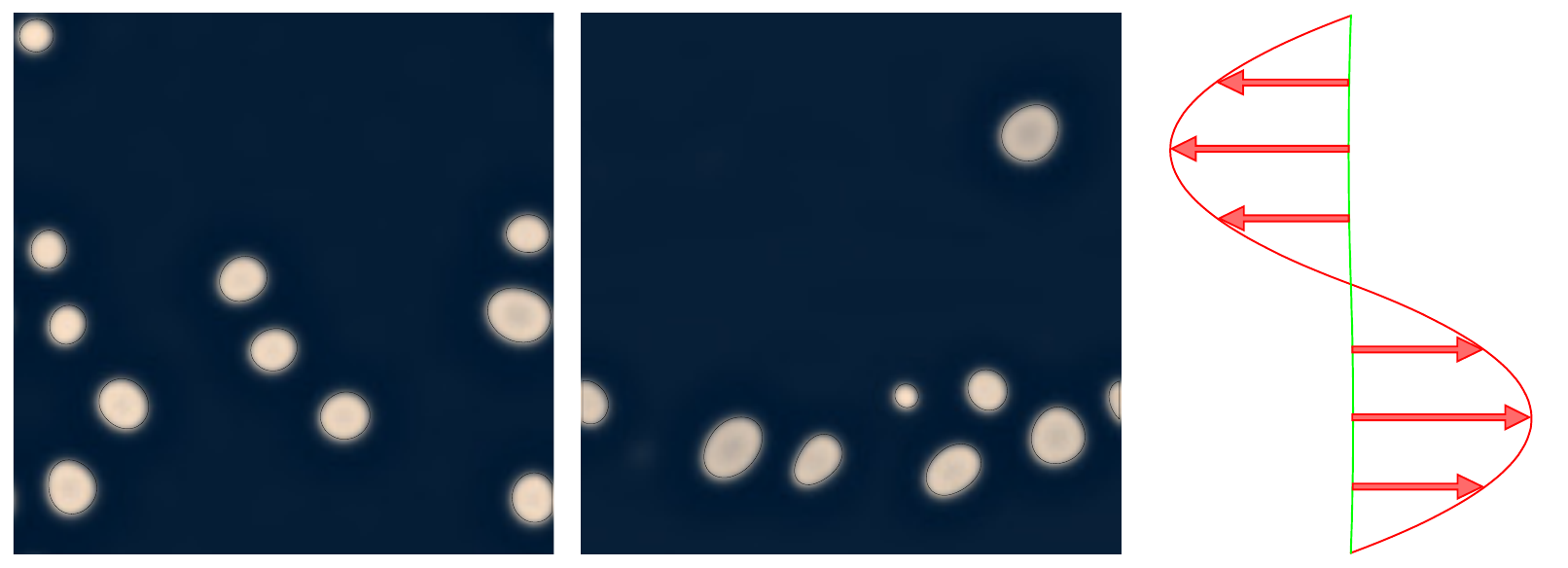}
\caption{Polymer density plot right after the nucleation stage at $t=700 \tau$ 
in droplet systems ($\chi_{BS}=0.75$) and force amplitude
$\fO = 1 \cdot 10^{-5} f^*$ (left snapshot) and 
$\fO = 8 \cdot 10^{-4} f^*$ (right snapshot). Corresponding flow
profile is shown on the right.}
\label{fig:075_nucleation}
\end{figure}

Next we examine the distribution of nucleation events across the channel (in
the $y$-direction perpendicular to the flow). \figref{fig:assembly_comp}
shows scatter plots of nucleation events in the plane of time vs.\
$y$-coordinate relative to the center of the Poiseuille flow (denoted $y=0$) for
different force amplitudes $\fO$ and our two choices of $\chi_{BS}$.  Here
nucleation events in the upper and lower regions of the simulation box with
opposite Poiseuille flows are shown together in one graph.  For small $\fO$,
nucleation events are distributed evenly across the channel.  For larger $\fO$,
nucleation preferably takes place in the area of lowest shear rate close to
$y=0$. \figref{fig:075_nucleation} shows examples of simulation snapshots
(polymer density plots) for weak and strong shear flow right after the
nucleation stage.  In the case of strong shear flow, the nuclei are close to
the center of the Poiseuille flow. In systems with weak shear flow, no such
preference can be observed.

\figref{fig:assembly_comp} also shows that nucleation events are more
focussed at the center of the flow in vesicle systems (left) than in 
droplet 
systems (right). Let us consider, for example, the histogram of nucleation
events in vesicle systems at force amplitude $\fO = 2 \cdot 10^{-4} f^*$, which
features a broad peak at the center and almost no counts at positions $|y| > 7
R_G$.  To reach a similar level of focussing in the micellar systems, one has
to increase the force amplitude by roughly a factor of four up to $ \fO \sim 8
\cdot 10^{-4} f^*$. Hence droplets can also assemble in regions of higher local
shear, whereas for vesicles this is unlikely.

Below we will see that the polymer \new{composition} profile in the
$y$-direction changes during the ripening stage and a double-peak structure
emerges at late times.  During the nucleation stage, this structure cannot yet
be seen.

\begin{figure}[t]
\centering		
\includegraphics[scale=1]{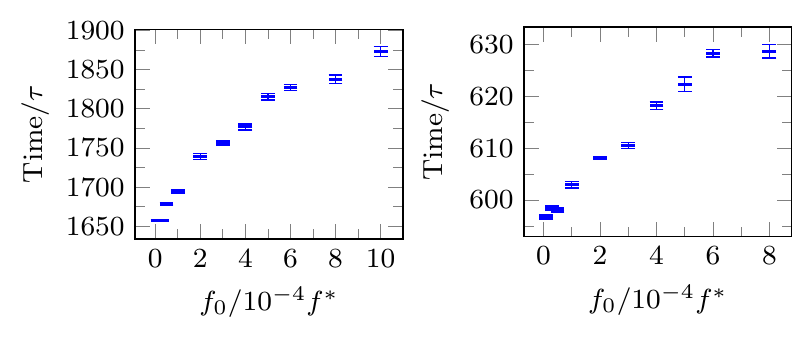}
\caption{Time of the nucleation stage as a function of 
force amplitude $\fO$ in units of $f^*$. 
Left: vesicle systems ($\chi_{BS}= -0.15$),  
right: droplet systems ($\chi_{BS}= 0.75$)}\label{fig:transition}	
\end{figure}

Next we examine the distribution of nucleation events in time. Already 
\figref{fig:assembly_comp} shows clearly that the nucleation stage is delayed if
the shear flow is increased. Furthermore, the width of the distribution of the
nucleation events in time is much broader in the vesicle systems than in the
droplet systems. To quantify this observation, we fit the distribution of
nucleation events in time by a Gaussian distribution. The fit matches the data
quite well, especially in systems with higher force amplitudes $\fO$, where we
get reduced $\chi^2$ values of the order 1-5. The first moment of the Gaussian
gives the characteristic time of the nucleation stage and is shown as a
function of $\fO$ in \figref{fig:transition}. Both in vesicle and
droplet systems, the nucleation stage is shifted to later times if shear
flow is applied.  The shift first increases with $\fO$ and then saturates at
large $\fO$, due to the fact that nucleation events are confined to the
low-shear center of the flow profile at such force amplitudes.

\begin{figure}[t]
\centering		
\includegraphics[scale=1]{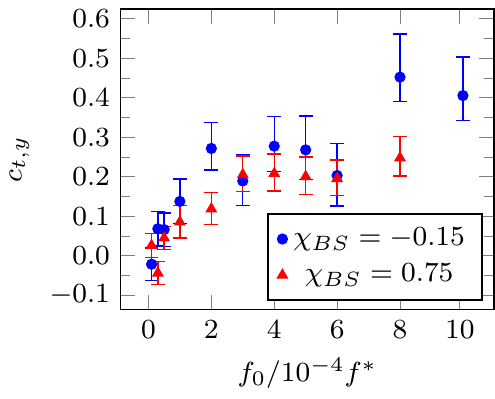}
\\
\caption{Correlation coefficient of time and $y$-coordinate of nucleation 
events as a function of force amplitude $\fO$ in units of $f^*$.
Blue: vesicle systems ($\chi_{BS}=-0.15$), 
red: droplet systems ($\chi_{BS}=0.75$)}\label{fig:assembly_corr}	
\end{figure}

Looking at \figref{fig:assembly_comp} more closely, it is apparent that the
time and $y$-coordinate of nucleation events are correlated.  For weak shear
flow, nucleation is homogeneous in $y$. For moderate shear flow, the scatter
plots have some resemblance with arrowheads pointing in the direction of
small times, i.e., nucleation events first take place close to the center of the
Poiseuille flow, and then become increasingly likely in areas with higher local
shear stress. To discuss these correlations more quantitatively, we calculate
the correlation coefficient of the coordinates $(t,y)$ of nucleation events,
defined as
\begin{equation}
    c_{t,y} =  \frac{\langle (t - \langle t \rangle )
                (|y| - \langle |y| \rangle ) \rangle}
               {\sqrt{\langle (t - \langle t \rangle )^2\rangle 
                 \langle (|y| - \langle |y| \rangle)^2 \rangle}}
\end{equation}
Here, $\langle ... \rangle$ denotes the statistical average. 
$c_{t,y}$ is equal to $1$ if $t$ and $y$ are perfectly correlated and
zero if there is no correlation. \figref{fig:assembly_corr} shows the
results as a function of force amplitude $\fO$. We find that the time and
position of nucleation events in Poiseuille flow are uncorrelated for small
shear flows, but they become correlated as the force amplitude $\fO$ increases.
In practice, this means that the distribution of nucleations gradually
broadens with time (see \figref{fig:assembly_comp}). The correlation
for vesicle systems and droplet systems is comparable.

\begin{figure}[t]
\centering		
\includegraphics[scale=1]{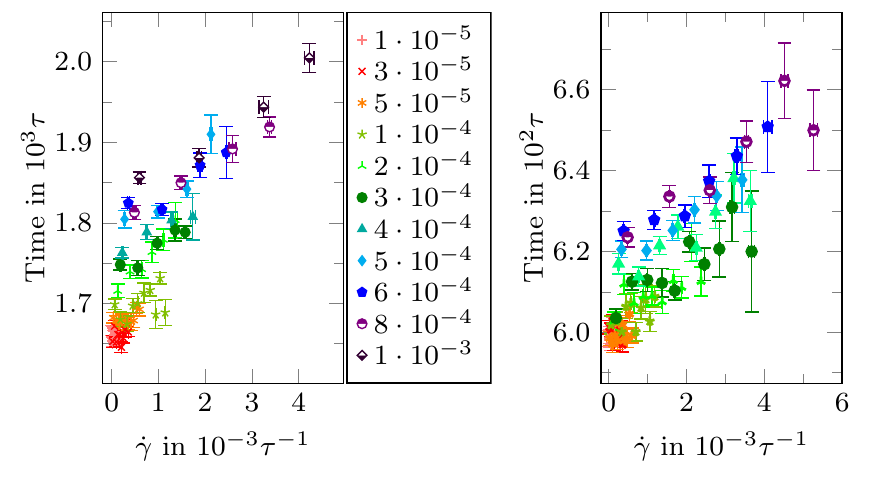}
\\
\caption{Mean time of nucleation events vs.\ local shear rate in Poiseuille
profiles for different driving force amplitudes $\fO$ as indicated
(in units of $f^*$) in vesicle systems with $\chi_{BS}=-0.15$ (left) and 
droplet systems with $\chi_{BS}=0.75$ (right).
}
\label{fig:shear_decay}
\end{figure}

We can also use our simulation data to investigate the relation between the
local shear rate of nucleation and the local delay time. To this end, we bin
the histograms in \figref{fig:assembly_comp} in the $y$-direction and
determine the mean delay time as a function of the local shear rate, for all
considered force amplitudes $\fO$, in the vesicle and droplet systems. The
results are combined in \figref{fig:shear_decay}. Especially for larger
shear rates and in the vesicle systems, the data roughly collapse on a single
almost straight line, i.e., the local nucleation time is roughly a linear
function of the local shear rate. At low shear rates, the data spread out due
to the effect of lateral polymer diffusion.  In the droplet systems where
the effect of local shear on the nucleation time is much weaker, the collapse
is less clear. Nevertheless, the local shear and the local nucleation time are
still strongly correlated.

These findings are consistent with experimental studies on spinodal
decomposition and structure formation in polymer mixtures in Couette flows
\cite{takebe1, takebe2, hashimoto1}.  Here, it was found that applying shear
flows has a similar effect on the length and time scales of spinodal
decomposition than shifting the spinodal line towards lower temperatures
\cite{hashimoto1}. If we adopt this interpretation, it follows that local shear
effectively shifts our system closer to the spinodal line, which in turn
increases the characteristic time scale of spinodal decomposition \cite{cahn1}
and hence the ''incubation'' time for nucleation \cite{he1}.

In sum, we find that shear flow significantly affects the droplet and
vesicle self-assembly in the nucleation stage. It affects both the time frame
and the preferred location of nucleation events. The central observation is
that nucleation is delayed in the presence of shear. This observation can
account for all findings reported here at a qualitative level: Due to the
shear-dependent delay, nucleation events are unevenly distributed in Poiseuille
flow. They first emerge in regions of low shear stress (the center of the flow
profile), and then gradually also populate regions with higher local stress. At
the same time, the existing nuclei grow by incorporating copolymers from
solution. The process stops when the remaining level of free copolymers is so
low that no further nucleation events take place. If the force amplitude is
strong, the nucleation stage is completed before any nucleation events have
taken place in the outer regions of the profile. As a result, strongly sheared
systems contain fewer nuclei than weakly sheared systems (
\figref{fig:max_part_density}) and their nuclei are concentrated around the
center of the profile (\figref{fig:assembly_comp}).  These effects are more
pronounced in vesicle systems than in droplet systems.

\subsection{Ripening stage: Evolution of particle number}
\label{sec:ripening}

\begin{figure}[b]
\centering		
\includegraphics[scale=0.74]{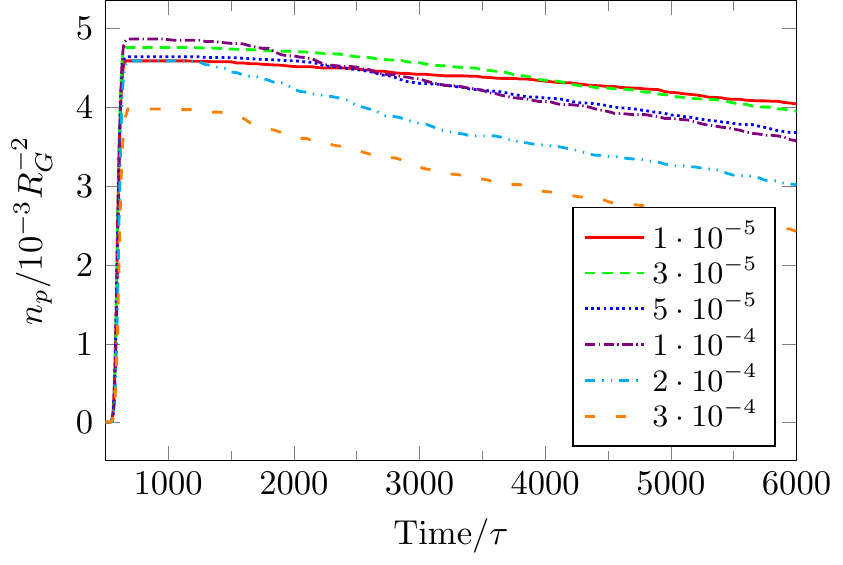}
\caption{Average particle number density in droplet systems ($\chi_{BS} = 0.75$)
as a function of time in units of $\tau$ for different force amplitudes
$\fO$ as indicated (units $f^*$).} 
\label{fig:150_075_part_l}	
\end{figure}

After the initial nucleation stage, the particle number remains constant or
decreases steadily. We will now focus on the evolution of the particle number
in this second, ''ripening'' stage.  We first consider the droplet
systems with $\chi_{BS} = 0.75$ (\figref{fig:150_075_part_l}).  In these
systems, a ripening process reminiscent of classical Ostwald ripening
takes place: Large particles tend to grow at the cost of smaller ones, since
they have an energetically more favorable surface to volume ratio.  The
equilibrium state in these systems (close systems without flows) is a single
phase separated droplet in solution.  Already in the absence of shear, the
system evolves slowly towards this final state. 

\begin{figure}[t]
\centering		
\includegraphics[scale=.5]{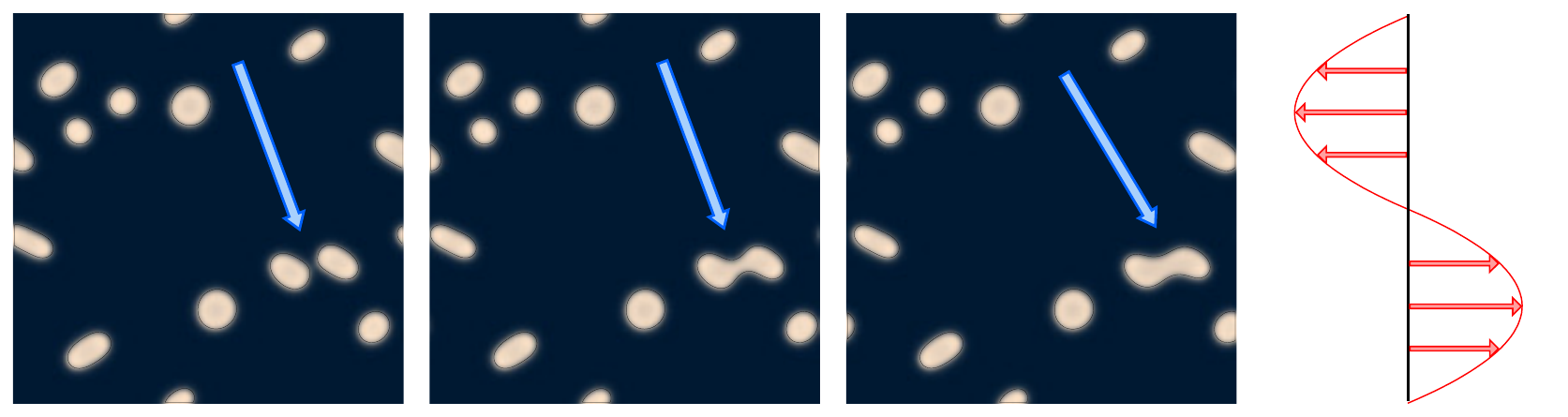}
\caption{
Series of snapshots in a droplet system ($\chi_{BS}=0.75$) with 
force amplitude $\fO=2 \cdot 10^{-4} f^*$ at times
$t=3350 \tau$, $t=3450 \tau$ and $t=3550 \tau$, showing the fusion 
of two particles with almost equal size (see blue arrows). Right panel
indicates the shape of the flow profile.}
\label{fusion_series}
\end{figure}

If shear flow is applied, the particle number decreases more rapidly.  A closer
inspection shows that this is not due to an acceleration of ripening, but due
to particle fusions. As discussed in \secref{sec:no_external_flow}, fusion
is suppressed in fluids at rest. Under the influence of shear, fusion events
become possible.  An example is shown in \figref{fusion_series}.

To analyze the ratio of particle fusions and particle dissolutions as a
function of shear strength, we must define criteria that distinguish between
two types of event -- particle traces getting lost due to particle dissolution
or due to fusion. This is done as follows: First, we exploit the fact that
only particles close to each other can fuse. Therefore, one criterion
for a fusion event is that two particles $i$ and $j$ with a distance less than
a threshold $d_{ij}$ must vanish at the same time. The threshold is chosen
$d_{ij} = 2.5(R_i + R_j)$, where $R_{i,j}$ is the radius of a spherical 
particle with the same polymer content as particle $i,j$, and the factor
$2.5$ accounts for the fact that particles may be deformed in shear flow.

In some rare cases it may happen that the sizes of two fusion partners are so
different, that only one of them vanishes and the other one remains nearly
unaffected. To distinguish between fusion and dissolution of particles in such
cases, we apply the second criterion that only those events are counted as
fusion, where the vanishing particles have an area larger than $3.3 \: R_G^2$.
If a particle has \new{an area} below this threshold before disappearing, we
assume that it has dissolved.

\begin{figure}[b]
\centering		
\includegraphics[scale=1]{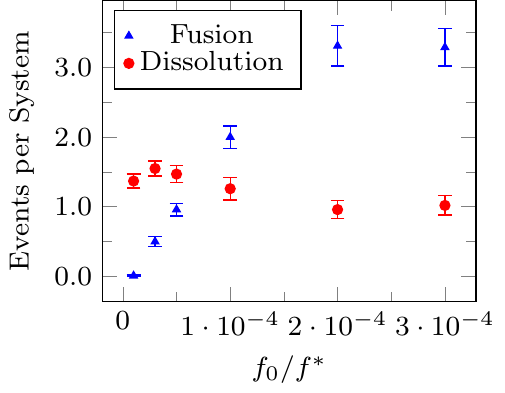}
\caption{Average number of fusions (blue) and average number of particle 
dissolutions per system (red) as function of $\fO$ (units $f^*$)
in droplet systems ($\chi_{BS}= 0.75$).
}\label{fig:dissol_fusion}	
\end{figure}

Using these criteria, we have determined the fusion and dissolution events
in droplet systems ($\chi_{BS}=0.75$) at low and moderate shear rates. The
results are shown in \figref{fig:dissol_fusion}. We find that the number
of particle dissolutions is almost independent of $\fO$. In contrast, the
number of particle fusions increases with $\fO$ and dominates for 
$\fO > 5 \cdot 10^{-5} f^*$. In systems with moderate shear, it
is 3-4 times larger than the number of particle dissolutions.

\begin{figure}[t]
\centering		
\includegraphics[scale=.74]{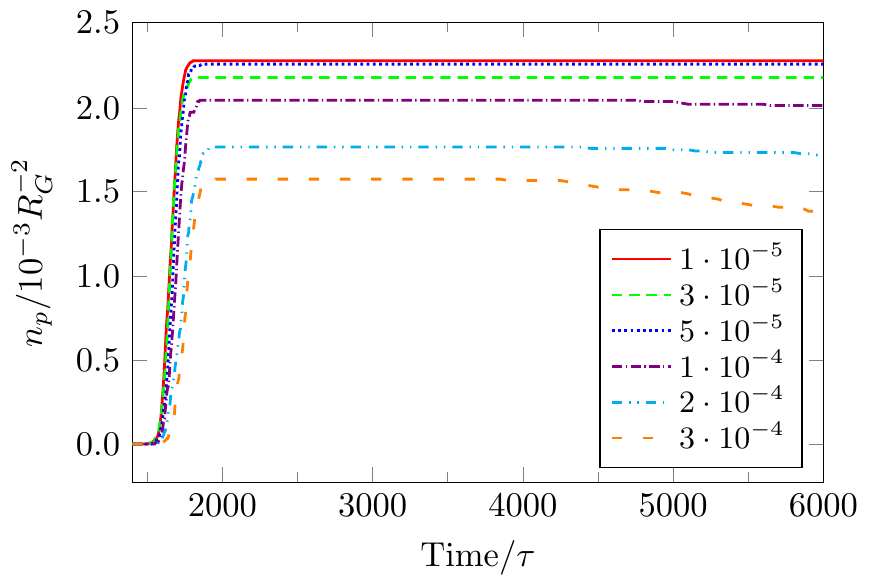}
\caption{Average particle number density in vesicle systems 
($\chi_{BS} = -0.15$) as a function of time in units of $\tau$ for 
different force amplitudes $\fO$ (units $f^*$).} 
\label{fig:150_015m_part_l}	
\end{figure} 

Next we examine the evolution of the number of particles in the vesicle
system ($\chi_{BS}=-0.15$). The results are shown in 
\figref{fig:150_015m_part_l}. In contrast to the droplet system, 
ripening is not observed in these systems (as already noted in Ref.\
\cite{he1}). In the absence of shear, the particle number does not change with
time. Under the influence of shear, it decreases, and this is the result of
particle fusions.

As discussed earlier and in Ref.\ \cite{he1}, fusion of micelle-shaped
droplets and vesicles is prevented in fluids at rest by the hydrophilic corona
surrounding the particles. Shear distorts the particles and disrupts the
corona, and as a result, fusion becomes possible. We find that even weak shear
flows can deform particles significantly. This will be discussed in the next
section. 

\subsection{Particle Shape}\label{sec:part_shape}

Next we consider the influence of shear on the shapes of particles. In our
systems with our model parameters, isolated particles at equilibrium tend to be
perfectly round.  In systems with more than one particle, the particles
influence and deform each other even without getting into contact.  In this
section, we study how the particle shapes change if the particles are exposed
to external shear flow.

The shape of a particle can be characterized by the tensor of gyration
\cite{theodorou1}. As the polymer distributions inside a particle are not
perfectly uniform in our case, we weight the distances between the different
lattice sites $i$ and $j$, which belong to a particle, with their polymer
densities $\Phi_P(\textbf{r}^i)$ and $\Phi_P(\textbf{r}^j)$: 
\begin{equation}\label{eq:gyration_tensor}
	G_{nm} = \frac{1}{2 {\cal N}^2} 
      \sum_i \sum_j \Phi_P(\textbf{r}^i) \Phi_P(\textbf{r}^j) 
             (r_n^i - r_n^j) (r_m^i - r_m^j)
\end{equation}
Here, $\cal N$ denotes the polymer content of the particle, and the sum 
runs over all lattice sites $i,j$ with $\Phi_P(\textbf{r}_{i,j}) \geq 0.5$.
By diagonalizing the tensor one obtains the eigenvalues $\lambda_-$ and
$\lambda_+$, from which the acircularity $c$ and the radius of gyration
$R$ can be derived,
\begin{equation}\label{asph}
	c = \lambda_+ - \lambda_-, \quad
	R = \sqrt{\lambda_+ + \lambda_-}
\end{equation}
Here, we will consider the relative acircularity, 
defined as \cite{theodorou1}:
\begin{equation}\label{rel_asph}
	c_{rel} = \frac{c^2}{R^4}
\end{equation} 
The relative acircularity is zero for perfectly circular particles
and one if the long axis is infinitely longer than the short axis. Thus, it
can be interpreted as the level of the particle deformation in
shear flow, and used to characterize both droplets and vesicles.

\begin{figure}[t]
\centering		
\includegraphics[scale=0.74]{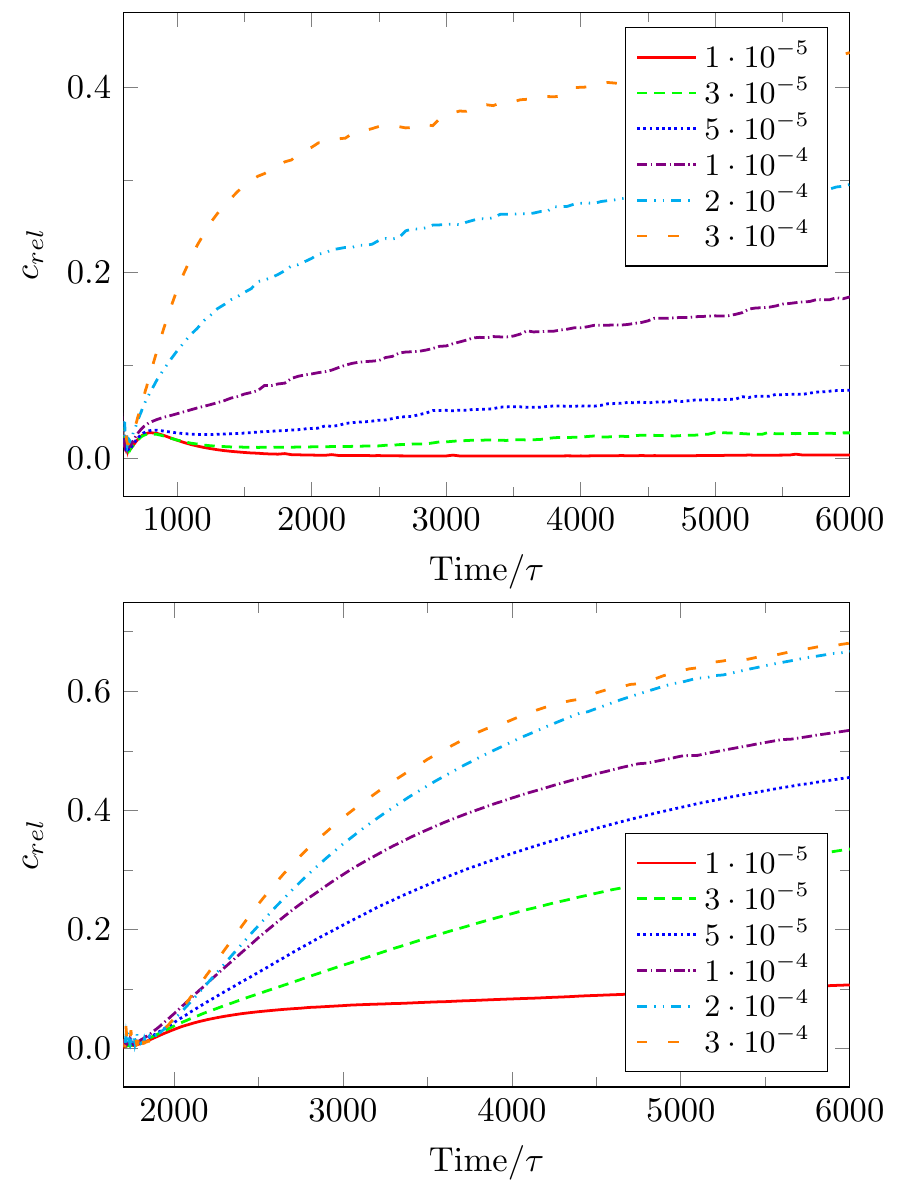}
\caption{Average relative acircularity of particles as a function of
time (units $\tau$) in droplet systems  with $\chi_{BS}=0.75$ (top) and 
vesicle systems with $\chi_{BS}=-0.15$ (bottom) for different force 
amplitudes $\fO$ (units $f^*$) as indicated.}
\label{fig:acirc_l}	
\end{figure}

In weakly sheared droplet systems (\figref{fig:acirc_l} top, $\fO < 3
\cdot 10^{-5} f^*$), the relative acircularity first increases and then
drops again. It reaches a maximum during the nucleation stage. This is because
freshly nucleated particles cannot grow isotropically if they are close to each
other, hence they deform slightly.  At later times, they gradually drift away
from each other and become more spherical, which lowers the relative
circularity again.
At the lowest force amplitude, $\fO = 1 \cdot 10^{-5} f^*$, the final
relative acircularity of the particles stays at a very low level. Thus we
can conclude that weak shear flow has no significant effect on the shape of
droplets in systems where the A- and B-monomers are both strongly
solvophobic. At moderate shear, the relative acircularity becomes more
pronounced and the shape of the curve changes (\figref{fig:acirc_l} top,
$\fO > 3\cdot 10^{-5} f^*$). After an initial relatively rapid increase
in the nucleation stage, $c_{rel}$ continues to grow more slowly at later
times. 

In the vesicle systems, shear flow is found to have a pronounced effect on
$c_{rel}$ of particles for all considered shear rates, and $c_{rel}$ keeps
increasing steadily at late times even in the system with lowest shear 
($\fO = 1 \cdot 10^{-5} f^*$). Hence particles in vesicle systems can be 
deformed much more easily than particles in droplet systems. We will now
analyze this in more detail.

\begin{figure}[t]
\centering		
\includegraphics[scale=.74]{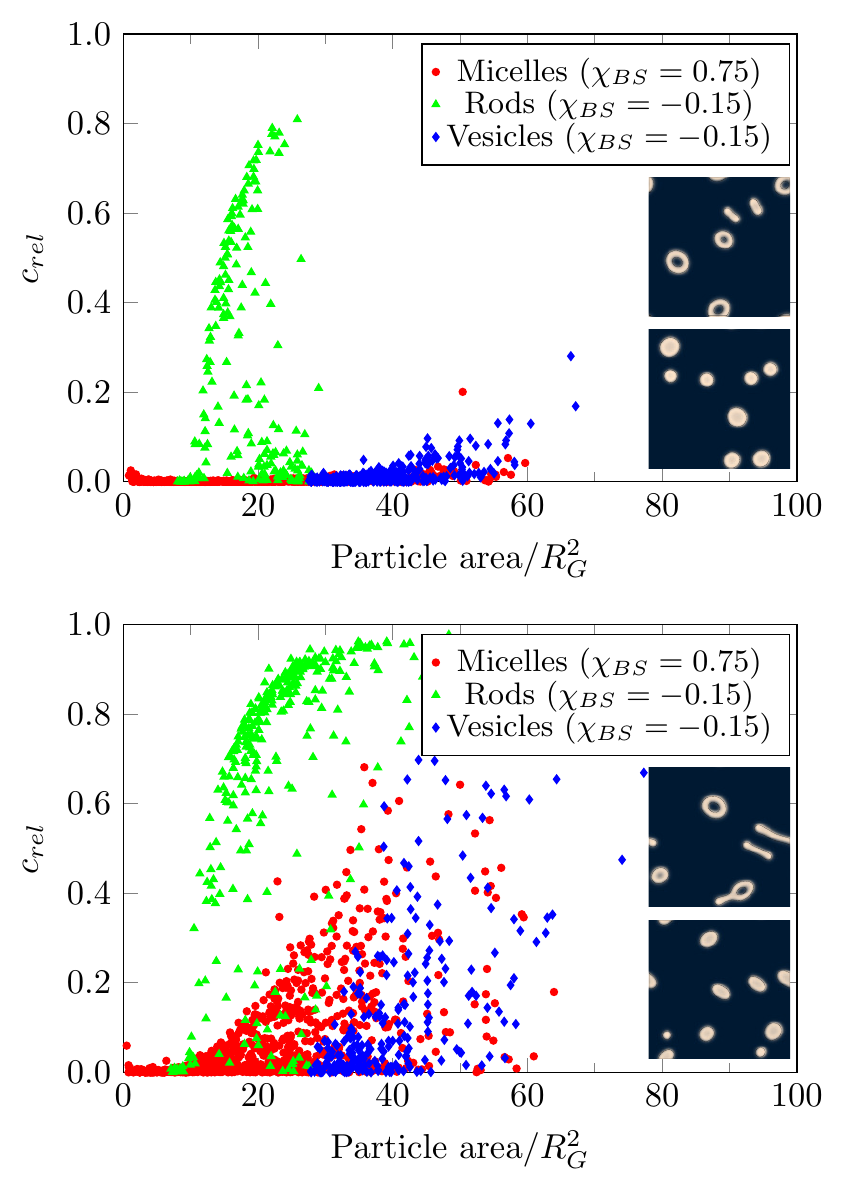}
\caption{
Scatter plot of particle characteristics in the plane of \new{particle area}
(units $R_g^2$) vs.\ relative acircularity in vesicle systems with
$\chi_{BS}=-0.15$ (green: rods, blue: vesicles) and droplet systems with
$\chi_{BS}=0.75$ (red: micelle-shaped droplets)  at time $t=6000 \tau$ for
force amplitude $\fO = 1 \cdot 10^{-5} f^*$ (top) and $\fO = 5 \cdot 10^{-5}
f^*$ (bottom). The inset shows snapshots from corresponding simulations at
$\chi_{BS}=-0.15$ (top) and $\chi_{BS}=0.75$ (bottom)
}
\label{fig:acirc_size_s05r}	
\end{figure}

One quantity that clearly influences the relative deformability of a particle
is its size. \new{Here we will examine the particle area, which we define} as
the area covered by a particle (interiors of hollow particles excluded), and
calculate it according to a procedure described in \appref{sec:particle_size}.
The average particle \new{area} is much larger in vesicle systems than in
droplet systems.  One might suspect that this is the main reason for their
higher deformability.  To investigate this possibility, we have constructed
scatter plots of particle size vs.\ relative particle acircularity for fixed
(late) simulation time and force amplitude $\fO$. Two examples are shown in
\figref{fig:acirc_size_s05r}.  Each symbol corresponds to one particle in
either a droplet or a vesicle system. In addition, we distinguish between
hollow and filled particles in the vesicle systems. For reasons that will
become clear below, we will refer to the latter ones as ''rods''.

In systems with small or moderate $\fO$, the regions in the
\new{area}-acircularity plane where certain particles exist and regions where
they apparently cannot exist are clearly separated. If particles are very
small, the range of accessible acircularities is generally very limited, i.e.,
they remain close to spherical. This observation is in good agreement with
early work on fluid droplets immersed in another fluid \cite{taylor1}. For
larger particle \new{areas}, the diagram displays a steep transition, beyond
which the acircularity limit is close to one. This limit is reached by a
special class of particles in the vesicle systems, which differ from regular
vesicles in that they do not enclose solvent (green triangles in
\figref{fig:acirc_size_s05r}), indicating that they correspond to elongated
micelles (see also the snapshots in \figref{fig:acirc_size_s05r}). In three
dimensions, they could correspond to either wormlike or disklike micelles.
Experimental observations suggest that shear flows with uniform shear rate can
induce shape transformations from vesicles into wormlike micelles
\cite{mendes1}.  Therefore, we will call these elongated structures ''rods''
hereafter. 

The other categories of particles (red spheres and blue diamonds) have much
lower acircularity.  A very small number of symbols lie outside the domain of
typically ''allowed'' acircularities. They correspond to particles that have
just emerged from a fusion event and are still highly non-circular. At later
times, they relax and become circular again. Interestingly, the accessible
range of acircularities for vesicles is smaller than that for droplets with
comparable \new{area}. Hence, contrary to expectations, we find that vesicles
show more resistance to deformations than micelle-shaped droplets.
Nevertheless, the total relative acircularity of particles is higher for
vesicle systems than for droplet systems (\figref{fig:acirc_l}) due to the
contribution of the rods.

\begin{figure}[t]
\centering		
\includegraphics[scale=.74]{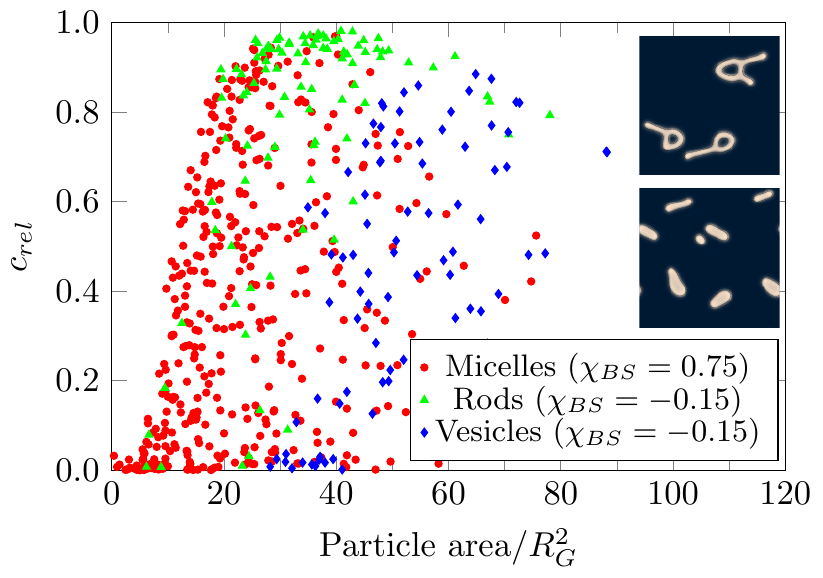}
\caption{
Same as \figref{fig:acirc_size_s05r} for force amplitude
$\fO = 3 \cdot 10^{-4} f^*$ at time  
$t = 3000 \tau$ for droplet systems ($\chi_{BS}=0.75$) and $t = 4500 \tau$
for vesicle systems ($\chi_{BS}=-0.15$).
}
\label{fig:acirc_size_s3r}	
\end{figure}

The acircularity limits for vesicles and micelle-shaped droplets are
found to depend strongly on the force amplitude $\fO$. If $\fO$ is very small,
as in \figref{fig:acirc_size_s05r} (top), all particles except the rods are
close to spherical. If $\fO$ is moderate, as in \figref{fig:acirc_size_s05r}
(bottom), the particles can deform more strongly. Moreover, vesicles may
develop ''fingers'' (see the snapshots in \figref{fig:acirc_size_s05r}
(bottom)). For large shear flows, i.e., large $\fO$, there are almost no
limitations on acircularity (\figref{fig:acirc_size_s3r}). Only very small
particles remain circular. 




We should note that \figref{fig:acirc_size_s3r} differs from 
\figref{fig:acirc_size_s05r} in that it does not show the size-acircularity
distribution at the end of a simulation, but at an earlier time where the
average acircularity still evolves strongly with time, especially in the
vesicle system. This is because at later times, more and more vesicles turn
into rod particles {\em via} an intermediate state of vesicles with fingers
(see inset of \figref{fig:acirc_size_s3r}), such that there are no
vesicles left for the analysis. The rod particles then simply maximize the
relative acircularity. Similar shape transformations of vesicles are also
observed at lower shear rates.

\begin{figure}[t]
\centering		
\includegraphics[scale=0.74]{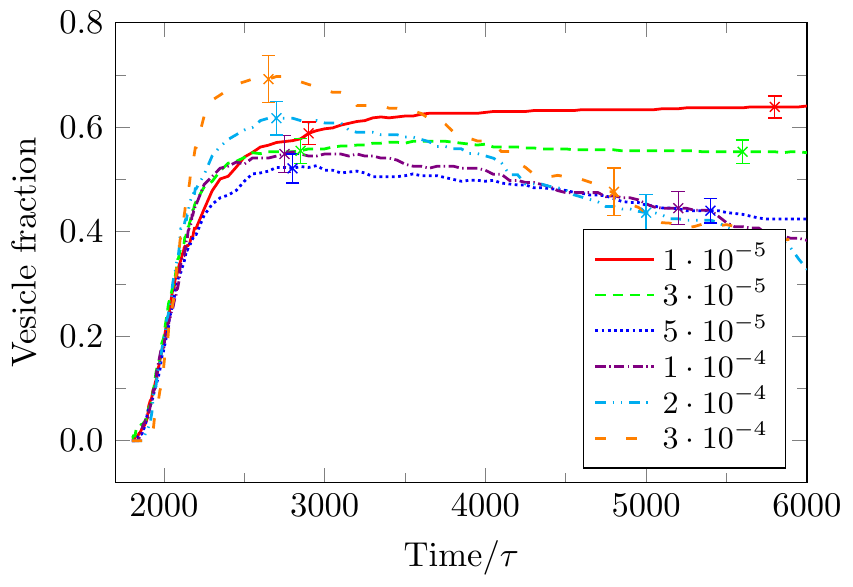}
\caption{Vesicle fraction in systems with $\chi_{BS}=-0.15$ vs.\ time
(in units of $\tau$) for different force amplitudes $\fO$ (in units
of $f^*$) as indicated. The total number of particles entering this
statistics is about $200$ for $\fO \geq 1 \cdot 10^{-4} f^*$ and 
$400$ for $\fO \leq 5 \cdot 10^{-5} f^*$. }
\label{fig:ves_015m}
\end{figure}

To analyze this more quantitatively, we will now focus on the vesicle systems
and investigate the fraction of vesicles with respect to the total number of
particles (vesicles and rods) as a function of time. The data are shown in
\figref{fig:ves_015m}. In weakly sheared systems ($\fO \le 3 \cdot 10^{-5}
f^*$), the fraction of vesicles in the system rises monotonically and reaches a
plateau at late times at around 0.6. A small increase of the shear rate shifts
the plateau to slightly lower values, but does not destroy it.  Once formed,
most vesicles hence tend to remain vesicular.  However, this is no longer true
in moderately or strongly sheared system. Here, the fraction of vesicles
reaches a peak shortly after the nucleation stage, whose height may even exceed
the value of 0.6 reached in the weakly sheared systems: Since the number of
nuclei is reduced in the presence of shear flow (see \secref{sec:nucleation}),
particles grow larger and are more likely to turn into vesicles. At later
times, the fraction of vesicles drops.  This is because the existing vesicles
first develop fingers and then eventually turn into rodlike particles. For even
more strongly sheared systems (data not shown), the height of the maximum
decreases and the rate with which vesicles disappear increases further.

Thus we conclude that in the vesicle system, the particle assembly in
Poiseuille flow proceeds in three stages: (i) Nuclei form. (ii) Nuclei
grow and turn into vesicles, much like in the closed system without flow
\cite{he1}. (iii) Vesicles may develop fingers which then grow and may
eventually transform the vesicle into a rod. The rate at which such
fingers appear increases with increasing shear rate.  However, fingering was
observed for all shear rates, even (rarely) in the most weakly sheared systems.
At the end of an infinitely long simulation, strongly sheared systems will
presumably only contain rods. At finite times, one has a mixture of droplets,
vesicles, and rods. 

\subsection{Characteristic relaxation times} \label{sec:relaxation}

\new{The deformability of particles in shear flow should depend on their 
relaxation time $\tau_d$, which sets the relevant mesoscopic timescale in 
the system. More specifically, we expect that shear flow starts to have
a significant effect on particles once the dimensionless shear rate,
$\tau_d \dot{\gamma}$, becomes of order unity. Assuming that $\tau_d$
increases with particle size, this would explain why larger particles
deform more easily than smaller particles. To test this assumption,}  
we will now examine the characteristic relaxation times $\tau_d$ of the
self-assembled particles in our systems. They were measured by taking
configurations of moderately sheared systems, stopping the flow in an instant,
and letting the particles relax. The relaxation of the acircularity with time
was then fitted to a single exponential, $c(t) = c_0 \exp(-t/\tau_d)$. 

\begin{figure}[t]
\centering		
\includegraphics[scale=1]{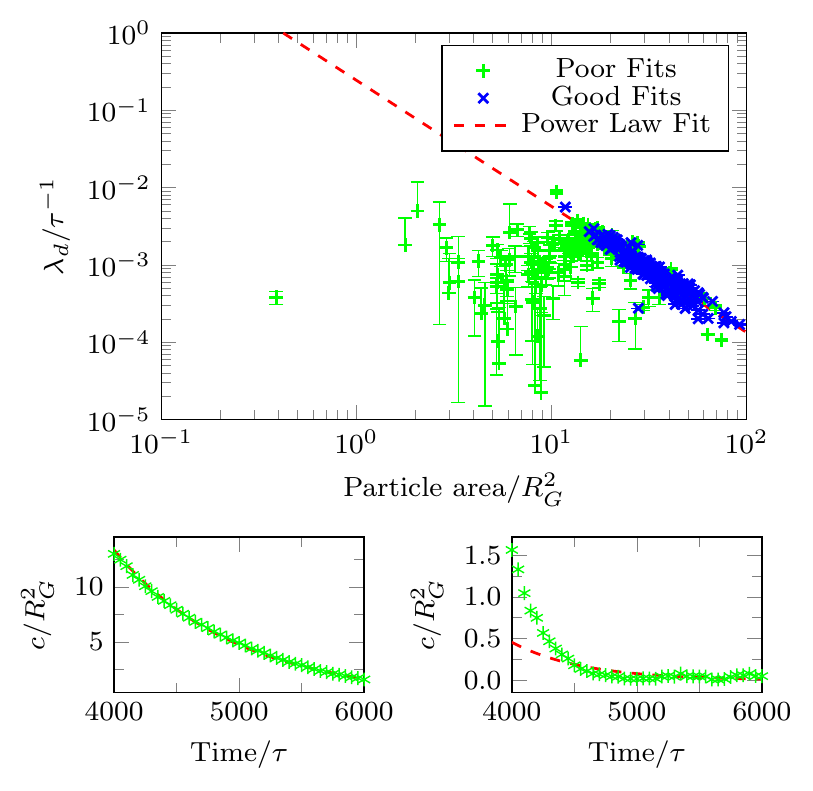}
\caption{Top: relaxation parameter $\lambda_d$ (in units of $1/\tau$)
in droplet systems ($\chi_{BS} = 0.75$)
obtained from exponential fits to the time evolution of the particle
acircularity after suddenly stopping a Poiseuille flow,
as a function of particle \new{area} (units $R_g^2$). 
Red: power law fit, blue: points used in the power law fit, 
green: points not used for the fit. 
Bottom: example for a good fit (left) and bad fit (right). Symbols show 
data, red lines show fitted exponential.
}
\label{fig:powerlaw_fit}	
\end{figure}

We begin with discussing the droplet systems with $\chi_{BS}=0.75$.  For
large particles, the data for $c(t)$ are mostly well-described by a single
exponential law. In some cases, the fit fails (see 
\figref{fig:powerlaw_fit}), in which case the data  are not included in the
further analysis.  Small particles generally show a more complex relaxation
behavior due to the fact that their size also varies with time and they
sometimes even dissolve.

Specifically, we only consider particles that exceed a minimum \new{area} and
whose relaxation times can be fitted well enough that the uncertainty of
$\lambda_d = 1/\tau_d$ in the exponential fit is less than 2 percent. A
selection of such points is shown in \figref{fig:powerlaw_fit} (top) for the
droplet system and an initial force amplitude of $\fO  = 3 \cdot 10^{-4} f^*$.
For comparison, the green symbols show the fit values for particles that do not
fulfill the selection criteria. The double logarithmic plot in
\figref{fig:powerlaw_fit} suggests that the relaxation time $\tau_d =
1/\lambda_d$ increases algebraically as a function of particle \new{area}.
Fitting the data to a power law of the form $\tau_d = \tau_d^0 (A/R_G^2)^b$
(where $A$ is the particle area), we obtain the fit parameters $\tau_d^0$ and
$b$ shown in \tabref{tab:tau_d}. The values for the relaxation time are almost
independent of $\fO$ (they increases slightly for larger $\fO$), and scale
approximately as $\tau_d \sim A^{3/2}$, or $\tau_d \sim R_d^3$, where $R_d \sim
\sqrt{A}$ is the equivalent particle radius. Hence, $\tau_d$ indeed increases
with $R_d$ as expected.  Inserting the data from \tabref{tab:tau_d} for
particles of radius $R_d \sim 5 R_G$, we find that the relaxation time should
be in the range of $\tau_d \sim 1000 \tau$. For such particles, the regime
\new{$\tau_d \dot{\gamma} \sim 1$} is reached at force amplitudes $\fO \sim
10^{-4} f^*$, in the regime that we call ''moderate''.

\begin{table}[b]
\caption{Fit parameters for $\tau_d(A) = \tau_d^0 \cdot A^{b}$ 
      for droplet systems ($\chi_{BS} = 0.75$)}\label{tab:tau_d}
\begin{tabular}{|l||c|c|c|}
 \hline 
	$\fO \: [f^*]$ 
            & $1 \cdot 10^{-4}$ & $2 \cdot 10^{-4}$ & $3 \cdot 10^{-4}$ \\
\hline
	$b$ & $1.50 \pm 0.06$ & $1.60\pm 0.05$ & $1.63 \pm 0.04$ \\
\hline
	$\tau_d^0 \: [\tau]$ 
    & $6.0 \pm 2.2$ & $4.6 \pm 1.3$ & $4.1 \pm 0.9 $ \\
\hline
	\end{tabular}
\end{table}

In vesicle systems the determination of a law for the relaxation time is 
much more difficult, since the two different types of particles, vesicles
and rods, show different behaviour. In addition, vesicles develop fingers,
which is an irreversible shape transformation. The dominant relaxation 
process for rods is the restoration of the equilibrium thickness, and for
deformed vesicles, the restoration of the circular shape. For vesicles with
fingers, one has a superposition of both. Due to the diversity of particle
shapes, the results for the relaxation of the acircularity $c$ (data not
shown) do not follow a clear trend, except that larger particles tend to
have longer relaxation times than smaller ones. The relaxation times of 
particles with size $R_d \sim 3-5 R_G$ in vesicle systems ($\chi_{BS}=-0.15$)
range from values around $\tau_d \sim 1000-4000 \tau$, which is comparable
to the relaxation times in droplet systems.

In the literature, the deformability of particles is often described in terms
of the so-called capillary number
\cite{vananroye1,kaoui1,guido1,kraus1,taylor1} \new{Ca, which depends on
the shear rate, the viscosity inside the particle, the radius, and
the interfacial tension. Here, the particles were so small that an 
interpretation in terms of Ca was not possible.}

\subsection{Lateral Migration of Polymeric Matter in Poiseuille Flow}
\label{sec:stdev}

Finally in this section, we discuss the distribution of polymeric matter in the
Poiseuille flow. Right after the nucleation stage, the polymer distribution
basically reflects the distribution of nucleation events discussed in
\secref{sec:nucleation}, and it has a single maximum in the region of lowest
shear. Later, the polymer particles redistribute within the flow profile. 

\begin{figure}[t]
\centering		
\includegraphics[scale=.74]{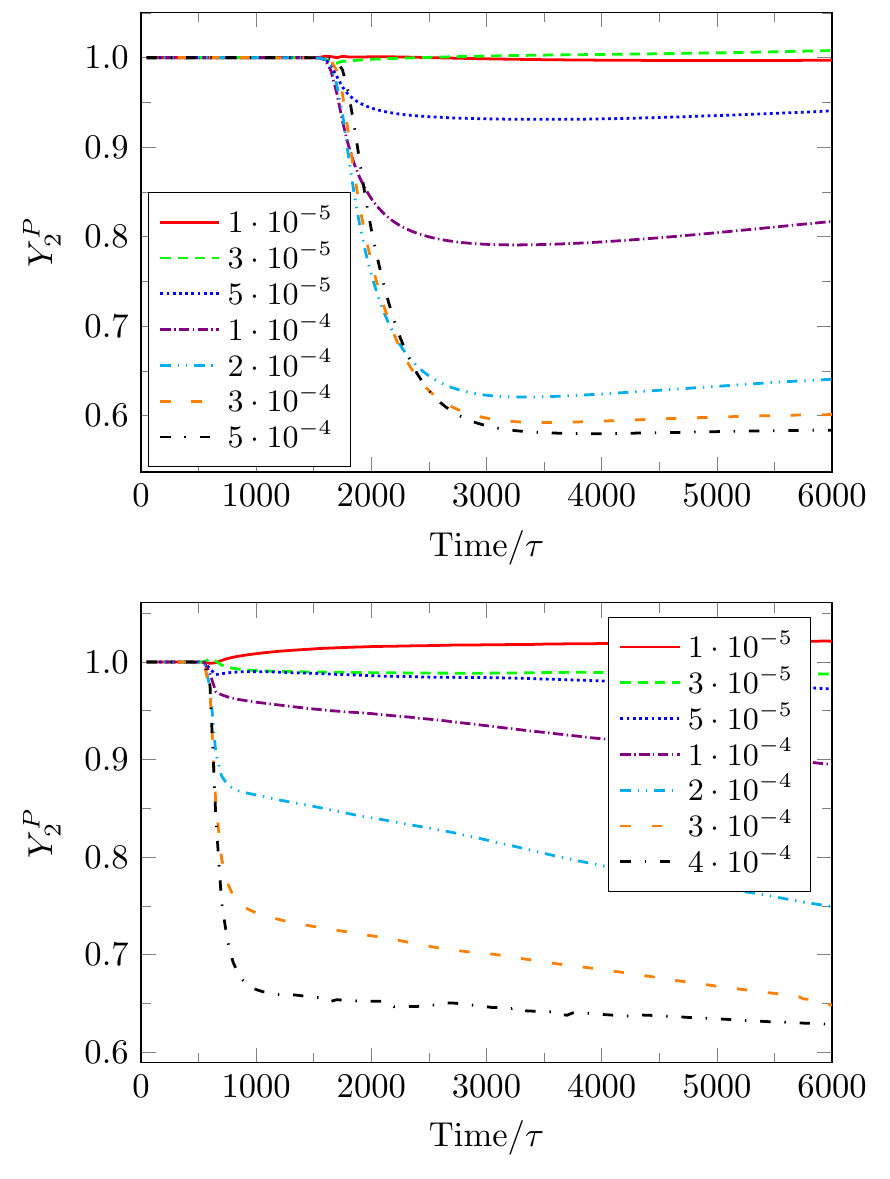}		
\caption{$Y_2^P$ as defined in Eq.\ (\ref{eq:width}) vs.\ time (in units 
of $\tau$) in vesicle systems (top: $\chi_{BS}=-0.15$) and droplet 
systems (bottom: $\chi_{BS}=0.75$) for different force amplitudes $\fO$ 
(units $f^*$) as indicated.}\label{fig:stdev}

\end{figure}

To analyze this effect, we introduce a quantity $Y_2^P$, which can be
interpreted as the normalized variance of the polymer distribution in the
direction perpendicular to the flow under the idealized assumption that the
polymers are symmetrically distributed around $y=0$.
\begin{equation}\label{eq:width}
    Y_2^P = \sqrt{\frac{\int_{-L_y/2}^{L_y/2} \mathrm{d}y \, y^2 
             \Phi_P (y)}{\bar{\Phi}_P \int_{-L_y/2}^{L_y/2} \mathrm{d}y \, y^2}},
\end{equation}
Here, $Y_2^P$ has been normalized such that a value of one corresponds to
a uniform distribution, values above one correspond to situations where the
polymeric matter preferably stays away from the center of the Poiseuille flow,
and values below one indicate that the polymeric matter is focussed near the
center.  Results for the time evolution of this quantity are shown in
\figref{fig:stdev}.  Both in vesicle and droplet systems, the polymeric
matter is distributed uniformly across the systems in the initial stage prior
to the first nucleation events. As nucleation sets in, $Y_2^P$ starts to
deviate from unity. 

We first examine the behavior for vesicle systems (\figref{fig:stdev}, top).
For weak shear rates, $Y_2^P$ stays close to one at all times.  For
moderate shear rates, it drops down rapidly, until it reaches a shallow
minimum. The level of the minimum decreases with increasing shear rates.  Its
position, $t \sim 2000-3000 \tau$, roughly corresponds to the time where the
vesicle fraction is largest according to \figref{fig:ves_015m}, suggesting that
the subsequent very slight increase of  $Y_2^P$ is associated with the
disruption of vesicles.  Finally, at strong shear rates,  $Y_2^P$
initially drops sharply and then saturates at a value around 0.6, which no
longer depends on the strength of the shear force.  In droplet systems
(\figref{fig:stdev}, bottom), the initial drop of $Y_2^P$ at intermediate
and strong shear rates is steeper than in the vesicle systems, almost
instantaneous, and it ends in a sharp crossover to a second regime where
$Y_2^P$ continues to decrease more slowly. A minimum is not encountered. 

\begin{figure}[t]
\centering		
\includegraphics[scale=1]{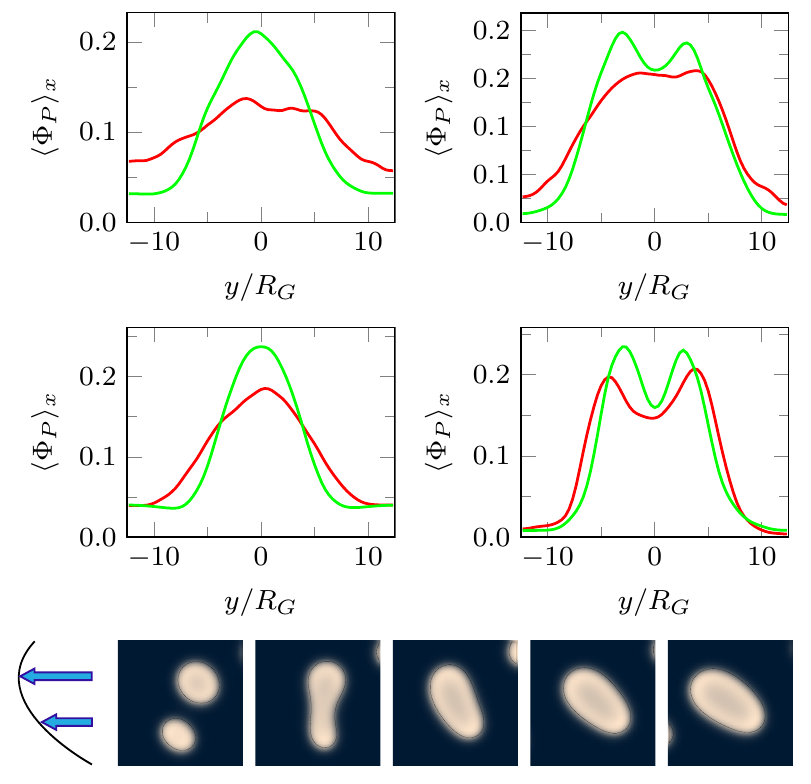}
\caption{Polymer \new{composition} profile in $y$ direction for vesicle systems
(green: $\chi_{BS}=-0.15$) and droplet systems (red: $\chi_{BS}=0.75$)
at force amplitude $\fO=2 \cdot 10^{-4} f^*$ (top) and 
$\fO =4 \cdot 10^{-4} f^*$ (middle) shortly after the nucleation 
stage (left: $t=700 \tau$ for droplet systems and
$t=2200 \tau$ for vesicle systems) and during the ripening stage 
(right: $t=6000 \tau$).
Bottom: Snapshots of a typical fusion process in a droplet system
($\chi_{BS}=0.75$) at $\fO =2 \cdot 10^{-4} f^*$ and times
$t = \{1500 \tau, 1900 \tau, 2800 \tau, 4000 \tau, 6000 \tau \}$} 
\label{fig:profile_s2r}	
\end{figure}

\figref{fig:profile_s2r} shows actual monomer density profiles shortly
after the nucleation stage (left) and at late times (right) for two different
shear rates (top and bottom) both for vesicle and droplet systems (green
and red lines).  At a qualitative level, the behavior of vesicle and
droplet systems is similar: Shortly after the nucleation stage, the
profiles feature a single peak close to the center of the profile. At later
times, a symmetric double peak structure emerges. 

We will now discuss the origin of this twin peak structure in more detail.  We
first focus on the droplet systems. Here, the monomer density
distribution at late times is governed by the interplay of ripening and
particle fusions. The ripening is driven by the competition of bulk and
surface energy, hence spherical particles at the center of the flow should grow
at the expense of more elongated particles in the outskirts of the profile.
This effect thus focusses matter to the center of the profile.  However, it
must be small, since the particle dissolution rate depends only weakly on the
shear in the system according to \figref{fig:dissol_fusion}.  On the other
hand, particle fusions typically drive matter away from the center of flow, as
demonstrated in \figref{fig:profile_s2r}, bottom. As discussed in
\secref{sec:ripening}, shear enables fusion events. Hence the center of mass of
two fusing particles will typically be located at a region of shear, $y \neq
0$, and the fusion event will drag matter to the periphery from the particle
which is closer to the center. 

In vesicle systems, the main mechanism responsible for the development of the
characteristic twin-peak structure is the fingering instability which
transforms the vesicles into rodlike particles as described in
\secref{sec:part_shape}. This mechanism leads to a net transfer of polymeric
matter from the vesicle in the flow maximum towards its "fingers" outside the
flow maximum. The alignment of the newly developed rods in the shear flow also
contributes to the focussing of polymeric matter at some distance to the 
flow maximum.

The lateral drift and the equilibrium position of particles or droplets in
Poiseuille flow has already been subject to many studies. In most of these
studies, the droplets were introduced as a separate fluid, which did not mix
with the solvent \cite{kaoui1, mortazavi1, doddi1, farutin1} and had a
conserved volume and in some cases even a conserved surface area. Therefore,
hydrodynamic boundaries played an important role. Even more importantly,
the systems under consideration also contained walls, which played a 
crucial role for the lateral migration behavior of the droplets. 

In our example we deliberately eliminated the effect of walls and did not
impose special hydrodynamic boundaries between particle and fluid. The
particle is considered a part of the fluid, it has the same viscosity,
and its only effect on the fluid dynamics is to impose surface forces generated
by the solvent-particle interfaces. This allows us to extract the pure
effect of the kinetics of self-assembly on the lateral distribution of
particles.  The observed lateral drifts are caused by the remodelling of
polymeric matter, which is driven by the free energy landscape and guided by
the non-local diffusive dynamics of the polymers. This kind of particle
deformation and reformation of polymeric matter has already been observed
before in simulations of polymer solutions or melts under shear flow
\cite{zvelindovsky1, zvelindovsky2} or in nanotubes \cite{feng1}.

\section{Discussion and Conclusions}
\label{sec:conclusions}

The two main messages of the present paper can be summarized as follows:

First, we have presented a new mesoscale simulation method for polymer
solutions, which couples a field-based dynamical model for diffusive polymer
motion with a nonlocal mobility function accounting for the chain connectivity
with a Lattice-Boltzmann scheme describing the hydrodynamic flows. It
extends a method proposed previously by Zhang et al.\ \cite{lzhang1}, which 
relied on a local dynamics assumption (monomers move independently). We have
shown that the new model reproduces experimental observations such as the
absence of vesicle fusions in closed systems. In previous simulations
based on local dynamics \cite{lzhang1}, fusions had been much too frequent.

The method allows us to study the dynamic evolution of inhomogeneous polymer
systems on length scales in the range of $\sim$ 10 nanometers to micrometers
and on time scales in the range of microseconds up to milliseconds.  For
example, in the present study, we can map the simulation units for length and
time, $R_G$ and $\tau$, to real SI-units by mapping the radius of gyration of
polymers (typically of order $R_G \sim 10 \mbox{nm}$) and the diffusion
constant of polymers in solution (typically of order $D_p \sim 10^{-6}
\mbox{cm}^2/\mbox{s}$), giving the time unit $\tau = R_G^2/N D_P \sim 0.1
\mu\mbox{s}$. Hence we simulated systems of size around $0.5$ micrometers over
a time of around $0.5$ milliseconds. 

Second, we have used our method to study the effect of shear flow on the
self-assembly of droplets or vesicles after a sudden quench from a homogeneous
copolymer solution. We have shown in \secref{sec:shear_flow} that shear flow
can be used to manipulate and control self-assembly in various ways. In the
following we recapitulate and discuss mainly those aspects that may turn out
relevant in particle design.

\paragraph*{(i) Particle number and particle size.} 
Poiseuille flow was found to affect the final number of particles in two ways.
First, shear flow increases the ''nucleation time'', i.e.\ , the characteristic
time when the first nuclei emerge after the quench. During the
narrow time window of the ''nucleation stage'' (which ends when most copolymers
from solution have been consumed), nucleation is therefore mostly restricted to
the central low-shear part of the Poiseuille flow profile. As a result, fewer
particles are nucleated in strong Poiseuille flow than in unsheared systems. A
second effect of shear flow effect which further reduces the particle number at
later times is the enhanced rate of particle fusions. Whereas particle fusions
are almost fully suppressed in the absence of shear, they become possible and
sometimes even dominate over particle dissolutions in the presence of shear.
Hence sheared systems contain fewer particles than closed systems.

This has consequences for the size of the self-assembled particles.  At given
copolymer volume fraction, the average particle size is inversely proportional
to the number of particles.  Hence the average size of self-assembled particles
should {\em increase} with the shear rate if self-assembly takes place in
Poiseuille flow, due to the fact that fewer particles are nucleated and they
merge more easily. 

It should be noted that in reality, the size of nanoparticles that are
assembled in microreactors is often found to {\em decrease} with increasing
flow rate \cite{thiermann_12}. The reason is that the quench rate in
microreactor setups is gradual and coupled to the flow rate -- for example, a
microfluidic device may be used to mix a component into a (co)polymer solution,
which induces (co)polymer aggregation.  In such cases, the flow rate controls
the particle size also {\em via} the quench rate \cite{nikoubashman_16,
kessler_16}, and as a result, particle sizes are smaller at larger flow rate. 

Hence the influence of shear on the particle size depends on the experimental
conditions. However, we can generally conclude that shear rates can be used to
control the size of particles {\em via} a variety of (sometimes competing)
mechanisms. 

\paragraph*{(ii) Shape transformations.} Under shear, both micelle-shaped
droplets and vesicles elongate. The elongation disappears if the flow is
stopped, and the particles become spherical again. However, irreversible shape
changes were also observed.  In particular, vesicles were found to develop
fingering instabilities and to turn into rod micelles at late times. Such
irreversible shape changes under shear could be used to design particle
structures that cannot be assembled under equilibrium conditions.

\paragraph*{(iii) Distribution of particles in Poiseuille flow.} We found
that self-assembled particles tend to be distributed in a twin peak structure
around the center of the Poiseuille flow profile. This can be used to design
particle distributions through inhomogeneous shear flows.

\paragraph*{Outlook.} It should be noted that all the results presented in the
present paper were obtained in two dimensional simulations. Some of the results
are presumably also valid in three dimensions. For example, we expect that the
emergence of nuclei after a sudden quench is still delayed in the
presence of shear, and that particle fusions are still facilitated by shear. In
other respect, the scenario in three dimensions can be expected to be quite
different. In particular, we can expect a much richer spectrum of irreversible
shape transitions under shear.  According to recent studies of vesicles in
Poiseuille flow one might expect parachutes or, in case the vesicles are not in
the flow maximum, slipper shapes \cite{farutin1,kaoui1}. In strong shear flows
it seems likely, that, according to \cite{zvelindovsky1, zvelindovsky2, cui1,
feng1}, fingers will develop and lead to long rods as in two dimensions. Hence
an extension of our study to three dimensions should be promising.

In real micro- and nanochannels, the presence of side walls also significantly
affects both the self-assembly and the flow effects. The present study was
deliberately set up to eliminate confinement effects and focus on the effect of
a spatially varying shear flow. In reality, the interplay of flow and
confinement should provide even more possibilities to design optimal flow
geometries for nanoparticle synthesis. Mesoscale simulation methods such
as the one proposed here should be useful to develop and assess such designs.

\acknowledgments

This work was supported by the VW foundation and by the German Science
foundation within TRR 146 (project C1). We wish to thank Burkhard D\"unweg,
Ryoichi Yamamoto, and Takashi Taniguchi for helpful discussions.  The
simulations were carried out at the supercomputing center MOGON of the Johannes
Gutenberg university. The configuration images were drawn with a
graph library developed by Alexander Wagner \cite{wagner1}.
\medskip

\begin{appendix}

\section{Rotation of the force field}\label{app:f_rotation}

In this appendix, we show that the force field $\textbf{f}_I(\textbf{r})$
defined by Eq.\ (\ref{eq:force_field}) is indeed irrotational within the
EPD approximation. Starting from Eq.\ (\ref{eq:force_field}), the rotation
of the force field can be calculated according to
\begin{align*}
	\nabla_{\textbf{r}} \times \textbf{f}_I(\textbf{r}) 
     &= - \nabla_{\textbf{r}} \times \sum_J \int \mathrm{d}\textbf{r}' 
    P_{IJ}(\textbf{r}, \textbf{r}') \nabla_{\textbf{r}'}
     \mu_J(\textbf{r}')\\
	&= \sum_J \int \mathrm{d}\textbf{r}' \nabla_{\textbf{r}'} 
      P_{IJ}(\textbf{r}, \textbf{r}')  \times \nabla_{\textbf{r}'} 
  \mu_J(\textbf{r}')
\end{align*}
Here, we have used the EPD approximation
$\nabla_{\textbf{r}}P_{IJ}(\textbf{r},\textbf{r}') = -
\nabla_{\textbf{r}'}P_{IJ}(\textbf{r},\textbf{r}')$ \cite{maurits1}. 
From \cite{kawasaki1} and \cite{maurits1} we know the
explicit form of the two-body correlator.
\begin{align*}
	&P_{IJ}(\textbf{r},\textbf{r}') 
	= \sum_{s,s'} P_{ss'}(\textbf{r},\textbf{r}') \delta_{sI}^K \delta_{s'J}^K \\
	&= \sum_{s,s'} n \langle \delta(\textbf{r}-\textbf{R}_s) \delta(\textbf{r}'-\textbf{R}_{s'}) \rangle \delta_{sI}^K \delta_{s'J}^K \\
	&=  \sum_{s,s'} n \int_{V^N} \mathrm{d}\textbf{R}_1...\mathrm{d}\textbf{R}_N \Psi \delta(\textbf{r}-\textbf{R}_s) \delta(\textbf{r}'-\textbf{R}_{s'}) \delta_{sI}^K \delta_{s'J}^K 
\end{align*}
Here, $s$ and $s'$ denote the index of the chain segment and $R_s$ is the
position of monomer $s$ and $\psi$ the single chain distribution function.
Hence we obtain
\begin{align*}
	&\nabla_{\textbf{r}} \times \textbf{f}_I(\textbf{r}) \\
	&= \sum_J \epsilon_{ijk} \int \mathrm{d}\textbf{r}' \big( \partial'_i P_{IJ}(\textbf{r}, \textbf{r}')\big) \big(\partial'_j \mu_J(\textbf{r}')\big) \hat{\textbf{e}}_k \delta_{sI}^K \delta_{s'J}^K\\
	&= \sum_J \epsilon_{ijk} \sum_{s,s'} \int_{V^N} \mathrm{d}\textbf{R}_1...\mathrm{d}\textbf{R}_N \Psi \delta(\textbf{r} - \textbf{R}_s) \\ 
	&\quad \times \int \mathrm{d}\textbf{r}' \big(\partial'_i \delta(\textbf{r}' - \textbf{R}_{s'})\big) 
	\big(\partial'_j \mu_J(\textbf{r}')\big) \delta_{sI}^K \delta_{s'J}^K \hat{\textbf{e}}_k
\end{align*}
Now, we can apply partial integration to obtain
\begin{align*}
	\nabla_{\textbf{r}} \times \textbf{f}_I(\textbf{r}) 
	&= -\sum_J \epsilon_{ijk} \sum_{s,s'} \int_{V^N} \mathrm{d}\textbf{R}_1...\mathrm{d}\textbf{R}_N \Psi \delta(\textbf{r} - \textbf{R}_s)\\ 
	& \quad \times \int \mathrm{d}\textbf{r}' \delta(\textbf{r}' - \textbf{R}_{s'}) \big(\partial'_i \partial'_j \mu_J(\textbf{r}')\big) \delta_{sI}^K \delta_{s'J}^K \hat{\textbf{e}}_k\\
	&= -\sum_J \int \mathrm{d}\textbf{r}' \epsilon_{ijk} P_{IJ}(\textbf{r},\textbf{r}') \big(\partial'_i \partial'_j \mu_J(\textbf{r}')\big)\\
	&= -\sum_J \int \mathrm{d}\textbf{r}' P_{IJ}(\textbf{r},\textbf{r}') 
            (\nabla \times \nabla \mu_J(\textbf{r}')) = \textbf{0}
\end{align*}

\section{Determination of the particle \new{area}}\label{sec:particle_size}
	
\begin{figure}[h]
	\centering		
	\includegraphics[scale=1.]{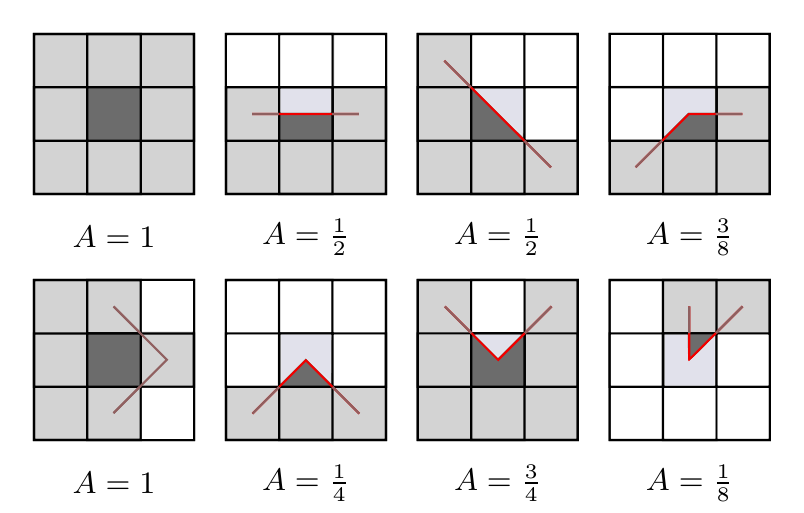}
	\caption{Weight factors $A$ with which a particle site
(the central site) contributes to the total particle area for
different local environments. Particle sites are shown in grey,
''empty'' sites (with polymer content $\Phi_P < 0.5$ are shown
in white.}
\label{fig:compound_area}	
\end{figure}

In \secref{sec:part_shape} we introduced the particle \new{area}. Here, we
explain briefly how it is determined from a \new{composition} distribution on a
grid.  First, we classify every lattice site with a polymer content of $\Phi_P
\geq 0.5$ as a particle site. The contribution of particle sites to the total
particle size depend on their local environment. Particle sites which are fully
surrounded by particle sites count as one, the others count partially as
illustrated in \figref{fig:compound_area}. We emphasize that the particle
\new{area} describes the area covered by the polymers of the particle, and not
its polymer content.

\section{Implementation of the simulation method}\label{sec:flowchart}

\begin{figure}[!h]
	\centering		
	\includegraphics[scale=1.]{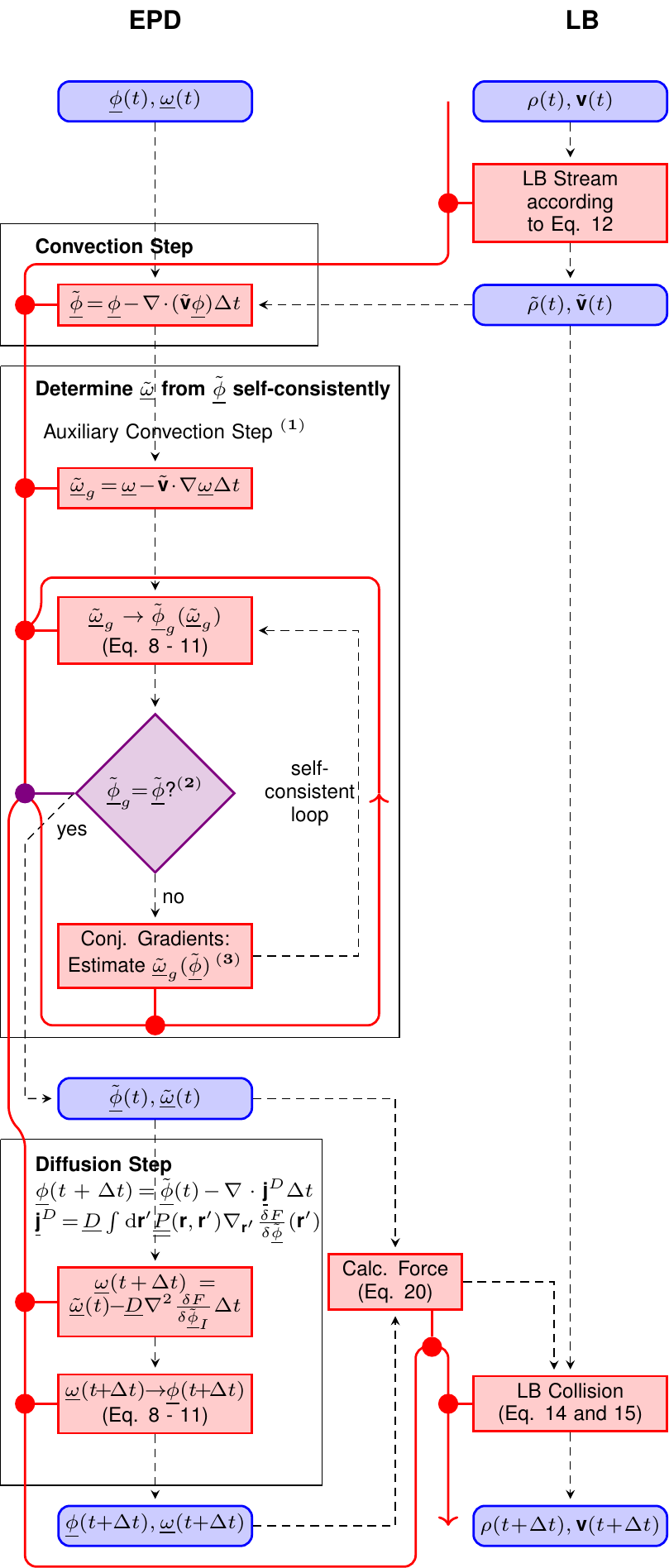}
	\caption{Flow chart diagram of the simulation program (one time step). 
See text for the comments (1,2,3) and for further explanation.}
\label{fig:flowchart}
\end{figure}

The simulation method combines two algorithms, which run in parallel and
pass information to each other when ever needed. The work flow of the code is
illustrated in the flow chart in \figref{fig:flowchart}.  Here we use the short
cut notation $\underline{\omega}$ and $\underline{\Phi}$ for the set of fields
$\omega_I, \Phi_I$ (with $I=A,B,S$), and $\underline{\underline{P}}$ for
$P_{IJ}$. The operations (red boxes) are divided into an EPD and a LB column.
The EPD column shows the evolution of the polymer-related fields
$\underline{\Phi}$ and $\underline{\omega}$, and the LB column the evolution of
the fluid-related fields $\rho$ (mass density) and $\mathbf{v}$ (fluid
velocity). The dashed black lines show the work flow within each columns and
mark the communication points.  The blue boxes show the status of the fields at
the beginning and end of a time step, and at selected intermediate states. The
continuous red arrow indicates the serial processing of the single operations
in our actual simulation program.

\new{Four aspects} in \figref{fig:flowchart} need to be explained in 
more detail:
\begin{itemize}

\item $\mathbf{^{(1)}}$ The auxiliary convection step of $\omega$ is 
introduced in order to avoid a pinning of polymeric structures
in cases where the fluid flow is so small that the changes of $\omega$ 
associated with the convection of $\Phi$ are below the accuracy threshold
of the iteration loop ($10^{-8}$ in our simulations, see below).

\item $\mathbf{^{(2)}}$ $\tilde{\underline{\phi}}_g$ and the desired
$\tilde{\underline{\phi}}$ are considered equal if  $\sqrt{\frac{1}{V} \int
\mathrm{d} \textbf{r} ( \tilde{\underline{\phi}}-
\tilde{\underline{\phi}}_g)^2} < N_{tr}$, where $N_{tr} = 1 \cdot 10^{-8}$ in
all simulation runs.  

\item $\mathbf{^{(3)}}$ To estimate $\tilde{\underline{\omega}_g}$, the
Fletcher-Reaves algorithm \cite{fletcher1} has been used.

\item \new{In the diffusion step, the current $\mathbf{j}^D$ is calculated
from the $\delta F/\underline{\delta \Phi}$ as evaluated at the intermediate
values of the composition field $\underline{\tilde{\phi}}$. Alternatively, one
could also evaluate $\delta F/\underline{\delta \Phi}$ at the original values
$\underline{\tilde{\phi}}(t)$. The resulting algorithm would have the same 
order (order one) as the present one. We have not compared the two 
algorithms. It will also be interesting to test more sophisticated, e.g., 
semi-implicit schemes.
}

\end{itemize}

\end{appendix}

\clearpage


\bibliography{paper_tc}

\clearpage
\begin{figure}[t]
\centering		

\includegraphics[scale=1.]{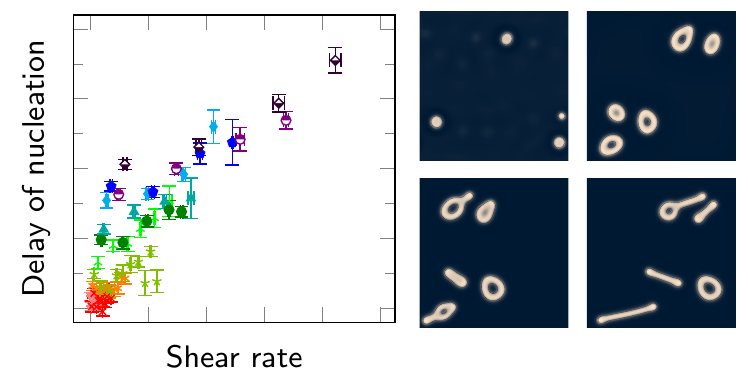}		
\caption{FOR TABLE OF CONTENTS ONLY. Title: Self-assembly of polymeric particles in Poiseuille flow: A hybrid Lattice Boltzmann / External Potential Dynamics simulation study. Authors: Johannes Heuser, G.\ J.\ Agur Sevink, Friederike Schmid}
\label{fig:toc}	
\end{figure}

\end{document}